\pdfoutput=1
\documentclass[aps,pre, reprint,groupedaddress, nofootinbib]{revtex4-1}
\usepackage{graphicx}
\usepackage{epstopdf}
\usepackage{physics}
\usepackage{amsmath,amssymb}
\usepackage[colorlinks=true, pdfstartview=FitV, linkcolor=blue, citecolor=blue, urlcolor=blue]{hyperref}
\usepackage[normalem]{ulem}
\begin{document}

\title{Satisfied-defect, unsatisfied-cooperate: An evolutionary dynamics of cooperation led by aspiration}
\author{Ik Soo Lim}
\email[]{i.s.lim@bangor.ac.uk}
\affiliation{School of Computer Science and Electrical Engineering, Bangor University, Bangor, Gwynedd LL57 1UT, United Kingdom
}

\author{Peter Wittek}
\affiliation{Rotman School of Management, University of Toronto, 105 St George St, Toronto ON M5S 3E6, Canada;}
\affiliation{Creative Destruction Lab, 105 St George St, Toronto ON M5S 3E6, Canada;}
\affiliation{Vector Institute for Artificial Intelligence, MaRS Centre, West Tower, Toronto ON M5G 1M1, Canada;}
\affiliation{Perimeter Institute for Theoretical Physics, Waterloo ON N2L 2Y5, Canada}

\date{\today}

\begin{abstract}
Evolutionary game theory has been widely used to study the evolution of cooperation in social dilemmas
where imitation-led strategy updates
are typically assumed.
However,
results of recent behavioral experiments are not compatible with the predictions based on the imitation dynamics,
casting doubts on the assumption of the imitation-led updates
and calling for alternative mechanisms of strategy updates.
An aspiration-led update 
is often considered as an alternative to the imitation-led one.
While details of update rules can have significant impacts on the evolutionary outcomes and many variations in imitation-led updates are thus studied,
however,
few variations exist in aspiration-led updates.
We introduce an aspiration-led update mechanism (``Satisfied-Defect, Unsatisfied-Cooperate'')
that is psychologically intuitive
and can yield a behavior richer than the conventional aspiration-led update does
in Prisoner's Dilemma games.
Using analytical and numerical methods,
we study
and link the stochastic dynamics of it in well-mixed finite populations
and the deterministic dynamics of infinite populations.
\end{abstract}


\maketitle

\section{Introduction}   

Explaining cooperation among selfish individuals in social dilemmas is an important problem
and has attracted much interest across  disciplines including physics 
\cite{Santos:2005ly, Szabo:2005ve, Sugiarto:2017bh, Stojkoski:2018qf}.
The Prisoner's Dilemma (PD) captures the problem of cooperation in the simplest and most challenging form  \cite{Veelen:2012aa}.        
Two individuals can choose between cooperation and defection. 
If one defects and the other cooperates, 
the defector gets a higher payoff than the cooperator does. 
They get a higher payoff if both cooperate than they do if both defect. 
Even though they would be better off if both cooperate,
individual (rational) reasoning leads to defection
 in  (one-shot) PD games.
This  illustrates a social dilemma due to the tension between the social optimum and individual interests.
In an evolutionary setting, 
the higher mean payoff of defectors implies more reproductive success (in genetic evolution)
and more imitation (in cultural evolution).
Cooperation is thus expected to perish.
However, cooperation is often observed in real-world social dilemmas.
        
Various mechanisms in the framework of the evolutionary game theory have been proposed to explain this apparent paradox. 
For the nongenetic evolution of cooperation,
it is typically assumed that successful strategies are spread by payoff-dependent imitation or social-learning \cite{Santos:2005ly, Szabo:2005ve, Sugiarto:2017bh, Stojkoski:2018qf}.
Payoff-dependent imitation means that an individual
first compares its payoff  and that of another individual,
and then copies the strategy of the other if the payoff of the latter is higher.
One of the main motivations behind the imitation-based strategy update is that
evolutionary game dynamics of the nongenetic evolution 
becomes formally equivalent to that of the genetic one \cite{Traulsen:2010zr};
similar mathematical models can describe both genetic and nongenetic evolution of cooperation.
Although it  is less applicable to lower animals
that lack cognition capabilities required for social learning,
the payoff-dependent imitation is considered adequate for humans 
and much theoretical work has been developed under this assumption
\cite{Santos:2005ly, Szabo:2005ve, Sugiarto:2017bh, Stojkoski:2018qf}.
However,
doubts have been cast  on the imitation-based update 
and  alternatives to it have been called for
\cite{Vilone:2014aa, Grujic:2014ys, cimini2014learning}.	
These are partly due to the recent behavioral experiments on PD games, 
which showed that
humans do not compare payoffs when updating their strategies
 \cite{Grujic:2014ys, Gracia-Lazaro:2012vn, Cimini:2015uq}.
 Indeed, it is often the case that individuals cannot even perceive the payoff of others
 in many real-world settings \cite{cimini2014learning}.
  
 An alternative to imitation-based social learning would be self-learning.
For instance, 
the aspiration-based mechanism of strategy updates
has been extensively investigated \cite{Amaral:2016zr, Liu:2016zr, Wu:2014aa, Patkowski:2009aa, Chen:2008hc, Szabo:2007tg, roca2011emergence, du2014aspiration, du2015aspiration, Posch:1999ij}.
According to the aspiration-based update,
individuals switch their strategy if the payoffs that they aspire are not met.
Unlike the imitation-based update,
it does not require any knowledge about the payoffs or strategies of others.
Hence,
it can be also applicable to the nongenetic evolution of cooperation in lower animals lacking cognitive capacities required for payoff-based imitation.
Indeed, aspiration-based strategy updates are often observed in studies of both animal and human behavioral ecology \cite{Bergen:2004ve, Galef:2008bh, Gruter:2011dq, Lopes:1999cr, Thompson:1988oq, Siegel:1957kl, Clay-Hamner:1975nx}.
While  imitation-based evolutionary dynamics yields cooperation to diminish,
 aspiration-based dynamics yields the emergence and sustainability of cooperation even in well-mixed infinite populations.

Since the details of the update rules can have significant influences on the emergence and stabilization of cooperation \cite{Szabo:2009fk},
it is well worth seeking alternative update rules led by aspiration.
To our knowledge,
however, few alternatives exist in aspiration dynamics.
We consider the whole space of aspiration-based update rules,
which includes the conventional aspiration-based update.
We formulate two psychologically intuitive properties that the desirable update rules should obey.       
Among all the rules,
only one satisfies both properties.
We analytically and numerically study the deterministic evolutionary dynamics and the stochastic dynamics  of the update rule as well as link them.
      
\section{Model Definition}

We consider the donation game version of PD games with two (pure) strategies of cooperation ($C$) and defection ($D$) in well-mixed populations.
A payoff matrix of the game is given by
\begin{equation}
\bordermatrix{
&C & D \cr
C &1 -\rho & -\rho \cr
D & 1 & 0
}
\end{equation}
where $\rho \in (0,1)$ denotes the cost of cooperation.
In a well-mixed population,
any pair of individuals play the game with the same probability.
In an infinite population, thus,
the mean payoffs of types $C$ and $D$ are
\begin{align}
\pi_{\scriptscriptstyle C}(x) &= x -\rho,\\
\pi_{\scriptscriptstyle D}(x) &= x
\end{align}
where $x \in [0,1]$ denotes the (relative) abundance or frequency of cooperators in the population.
In a finite population,
the mean payoffs are given by
\begin{align}
\pi_{\scriptscriptstyle C}(i) &= \frac{i-1}{N-1} -\rho,
\label{eq_payoff_c}\\
\pi_{\scriptscriptstyle D}(i) &= \frac{i}{N-1}
\label{eq_payoff_d}
\end{align}
where $N$ denotes the population size and $i$ the number of cooperators.

\section{Aspiration-based Strategy Update}

We consider the aspiration-based strategy updates
where the aspiration level is the same for all individuals of a population.
The conventional aspiration-led update 
can be stated as follows;
if one's payoff is higher than (or equal to) the aspiration level (i.e.\,$\pi \ge A$), then keep the current strategy
and  otherwise ($\pi <A$), switch it to the other strategy
\cite{Amaral:2016zr, Chen:2008hc, Szabo:2007tg, roca2011emergence, du2014aspiration, du2015aspiration, Posch:1999ij}.
We term this update rule `Satisfied-Stay, Unsatisfied-Shift' (SSUS),
which is a special case of the reinforcement learning \cite{Patkowski:2015uq}.
This deterministic rule is often relaxed to be stochastic, 
which reflects perception errors as well as other uncertainties
and drives the probabilistic change of the population composition.
To model the stochastic switching of the strategy,
the following probability functions based on Fermi functions are often used
\begin{align}
q_{\scriptscriptstyle D\rightarrow C}(\pi_{\scriptscriptstyle D},A) &= \frac{1}{1+\exp[-\beta(A-\pi_{\scriptscriptstyle D})]}, \label{eq_SSUS_d2c} \\
q_{\scriptscriptstyle C\rightarrow D}(\pi_{\scriptscriptstyle C},A) &= \frac{1}{1+\exp[-\beta(A-\pi_{\scriptscriptstyle C})]}
\label{eq_SSUS_c2d}
\end{align}
where
$q_{\scriptscriptstyle D\rightarrow C}$ denotes the probability for switching defection to cooperation,
$q_{\scriptscriptstyle C\rightarrow D}$ the probability for cooperation to defection,
 $\beta$ the selection intensity, 
 and $A$ the aspiration level \cite{Hauert:2005ff}.
The lower payoff $\pi$ than aspiration $A$, the more likely to switch the strategy;
the higher payoff $\pi$ than aspiration $A$, the less likely to switch the strategy.
As $\beta \rightarrow \infty$, the deterministic update rule is recovered.

\section{Satisfied-Defect, Unsatisfied-Cooperate}

\subsection{Space of strategy update rules}
In order to derive an update rule,
we explore the whole space of (deterministic) update rules led by aspiration with two strategies $C$ and $D$.
We can encode an aspiration-based update rule as  a finite state automaton
that has  two states of $C$ and $D$ with transitions between them being conditioned on whether  $\pi \ge A$ or not.
The encoding $\left(\mathcal{S}_{C+},\mathcal{S}_{C-},\mathcal{S}_{D+},\mathcal{S}_{D-}\right)$ specifies the strategy to be taken $\mathcal{S}_{C+}$ and $\mathcal{S}_{D+}$ given $\pi \ge A$ and the current strategy of $C$ and $D$, respectively.
It also specifies the strategy to be taken  $\mathcal{S}_{C-}$ and $\mathcal{S}_{D-}$ given $\pi < A$ and the current strategy of $C$ and $D$, respectively.
For instance, SSUS is encoded as $\left(C, D, D, C\right)$ (Table\,\ref{tab_automata}).  
\begin{table*}[th]
\caption{An encoding scheme of an aspiration-based update rule as a finite state automaton that describes transitions between two states of $C$ and $D$, conditioned on $\pi \ge A$ or not.
SSUS is encoded as $\left(C, D, D, C\right)$.
}
\begin{center}
\begin{tabular}{c|cc}
 & $\pi \ge A$ & $\pi < A$ \\
 \hline
$C$ & $\mathcal{S}_{C+}$ & $\mathcal{S}_{C-}$ \\
$D$ & $\mathcal{S}_{D+}$ & $\mathcal{S}_{D-}$ \\
\end{tabular}
$\Longleftrightarrow$
$\left(\mathcal{S}_{C+},\mathcal{S}_{C-},\mathcal{S}_{D+},\mathcal{S}_{D-}\right)$,
\hspace{0.2cm}
\begin{tabular}{c|cc}
 & $\pi \ge A$ & $\pi < A$ \\
 \hline
$C$ & $C$ & $D$ \\
$D$ & $D$ & $C$ \\
\end{tabular}
$\Longleftrightarrow$
$\left(C, D, D, C\right)$
\end{center}
\label{tab_automata}
\end{table*}
In total, there are $2^4$ finite state automata of this form.
We present two properties that a desirable update rule should obey,
each of which is psychologically intuitive.

The first property is conditional switching.
The strategy to be taken when `satisfied' (i.e.\,$\pi \ge A$) should differ from the one
when `unsatisfied' (i.e.\,$\pi <A$).
This property excludes those rules which yield the same strategy regardless of `satisfied' or not;
e.g.\,the rule of cooperate-no-matter $(C,C,C,C)$.
Among the 16 update rules,
four rules satisfy the property of the conditional switching, including SSUS (Table \ref{tab_automata_conditional}).
\begin{table}[t]
\caption{Four update rules that satisfy the conditional switching property.
}
\begin{center}
\begin{tabular}{cccc}
\hline\hline
$\left(C,D,C,D\right)$, & $(C,D,D,C)$, & $(D,C,C,D)$, &  $(D,C,D,C)$
\\ \hline\hline
\end{tabular}
\end{center}
\label{tab_automata_conditional}
\end{table}		
	
The second property is the selfishness or cost minimization while in satisfaction.
For the same outcome of `satisfied' (i.e.\,$\pi \ge A$), 
we assume that individuals prefer defection to cooperation since the latter incurs  a cost.
For a satisfied cooperator ($\pi_{\scriptscriptstyle C} \ge A$),
switching to defection makes sense in
that the aspiration 
is still expected to be met after the switching, but without the cost of cooperation
since $\pi_{\scriptscriptstyle D} >\pi_{\scriptscriptstyle C} \ge A$.
That is to say, 
one defects when the aspiration is met.
There are four update rules
which satisfy the cost-minimization property (Table \ref{tab_automata_cost_min}).
\begin{table}[t]
\caption{Four update rules that satisfy the cost-minimization property.
}
\begin{center}
\begin{tabular}{cccc}
\hline\hline
$(D,C,D,D)$, & $(D,D,D,C)$, & $\left(D,D,D,D\right)$, & $(D,C,D,C)$.
\\ \hline\hline
\end{tabular} 
\end{center}
\label{tab_automata_cost_min}
\end{table}

Among the 16 finite state automaton, 
there is only one rule that meets both of the properties,
which is $(D,C,D,C)$.
We name the \,$(D,C,D,C)$ rule `Satisfied-Defect, Unsatisfied-Cooperate' (SDUC).
The SDUC rule specifies defection to be taken when the aspiration is met ($\pi \ge A$) and cooperation to be taken when the aspiration is not met ($\pi <A$).
The switching probabilities of  SDUC are given by
\begin{align}
q_{\scriptscriptstyle D\rightarrow C}(\pi_{\scriptscriptstyle D},A) &= \frac{1}{1+\exp[-\beta(A-\pi_{\scriptscriptstyle D})]}, \label{eq_SDUC_d2c}\\
q_{\scriptscriptstyle C\rightarrow D}(\pi_{\scriptscriptstyle C},A) &= \frac{1}{1+\exp[-\beta(\pi_{\scriptscriptstyle C} -A)]}.
\label{eq_SDUC_c2d}
\end{align}
Note that it only differs in $q_{\scriptscriptstyle C\rightarrow D}$,
compared to those of SSUS [Eqs.\,\eqref{eq_SSUS_d2c} and \eqref{eq_SSUS_c2d}].
In the next sections,
we analyze the evolutionary dynamics of SDUC in infinite and finite populations
as well as  compare them with those of SSUS.

\section{Infinite Populations}

For a well-mixed infinite population,
the mean-field equation of deterministic evolutionary dynamics  is given by
\begin{equation}
\dv{x}{t} = (1-x)q_{\scriptscriptstyle D \rightarrow C} -x q_{\scriptscriptstyle C \rightarrow D}
\label{eq_mean_field}
\end{equation}
where the first term on the right-hand side captures the inflow of individuals switching to cooperation, 
and the second one, the outflow of those switching from cooperation to defection.
At $\dv*{x}{t}=0$,
we have an equilibrium frequency $x^*$ of cooperators,
which satisfies
\begin{equation}
x^* = \frac{q_{\scriptscriptstyle D \rightarrow C}}{q_{\scriptscriptstyle D \rightarrow C} +q_{\scriptscriptstyle C \rightarrow D}}.
\end{equation}
Under weak selection $0 <\beta \ll 1$,
we can approximate the equilibrium by
\begin{equation}
x^* 
\approx \frac{q_{\scriptscriptstyle D \rightarrow C}}{q_{\scriptscriptstyle D \rightarrow C} +q_{\scriptscriptstyle C \rightarrow D}}  \bigg\vert_{\beta=0} +\pdv{}{\beta}\left(\frac{q_{\scriptscriptstyle D \rightarrow C}}{q_{\scriptscriptstyle D \rightarrow C} +q_{\scriptscriptstyle C \rightarrow D}}\right)  \bigg\vert_{\beta=0}\beta.
\end{equation}
Prior to SDUC,
we start with the analysis of SSUS.
Although there already exists an analysis of SSUS in an infinite population,
it is limited only to the deterministic update, corresponding to the strong selection $\beta \rightarrow \infty$ \cite{Posch:1999ij}.
Our analytical study is based on a stochastic update or weak selection 
and we numerically study the strong selection cases as well.

\subsection{$x^*$ of SSUS}
\begin{figure*}[t]      
\begin{center}    
\begin{tabular}{cc} 
 \includegraphics[width=0.43\textwidth]{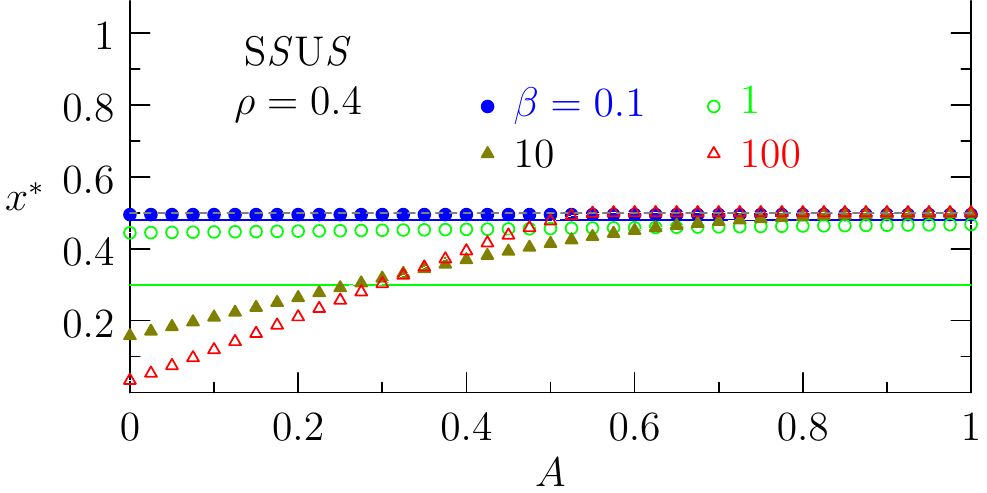}        &        
\includegraphics[width=0.43\textwidth]{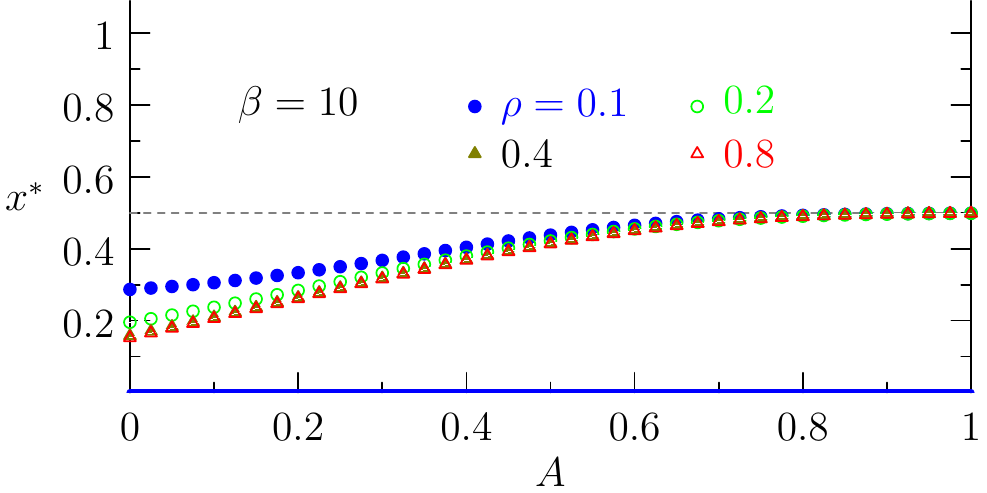}        \\
 \includegraphics[width=0.43\textwidth]{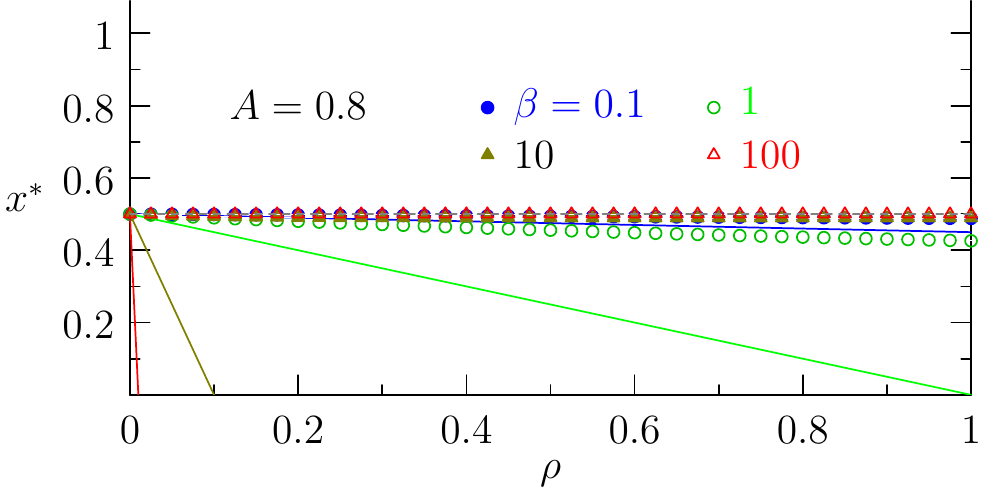}     &      
\includegraphics[width=0.43\textwidth]{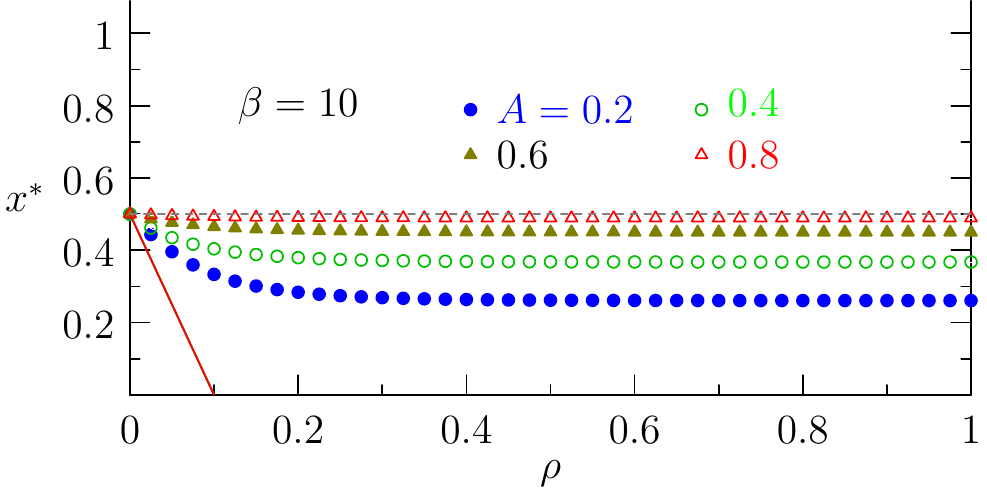}   \\
 \includegraphics[width=0.43\textwidth]{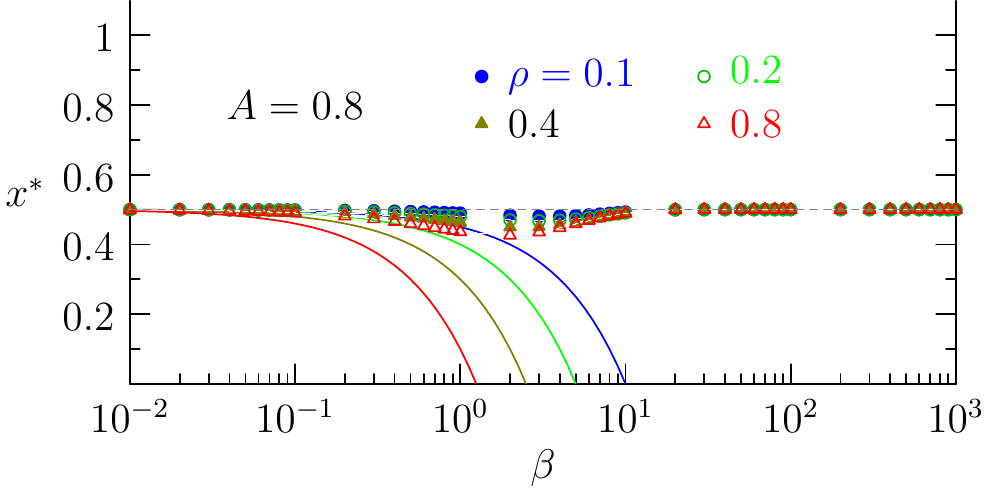}   &           
\includegraphics[width=0.43\textwidth]{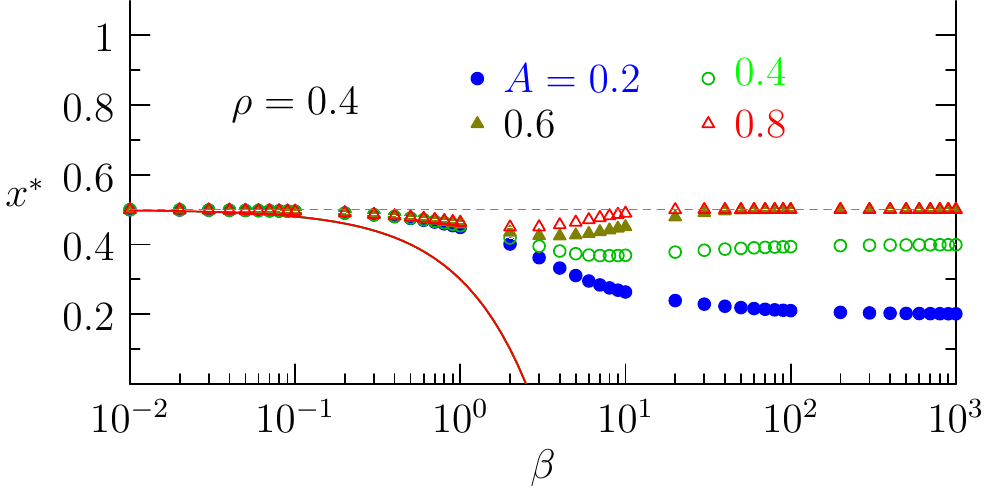}   \\
\end{tabular}     
\end{center}    
\caption{ 
The equilibrium $x^*$ of SSUS vs.\,aspiration level $A$,
cost $\rho$, and selection strength $\beta$.
The circles and triangles indicate the equilibrium obtained by numerically solving Eq.\,\eqref{eq_ssus_equilibrium}.
The solid lines indicates the analytical approximation of $x^*$ by Eq.\,\eqref{eq_ssus_equilibrium_approx},
which works well only for weak selection.
	The dashed line indicates $x^* =1/2$.
}   
\label{fig_equilibrium_SSUS} 
\end{figure*}

Under the SSUS rule,
at equilibrium,
we get
\begin{equation}
x^* 
=\frac{1+\exp[-\beta(A -\pi_{\scriptscriptstyle C})]}{2 +\exp[-\beta(A -\pi_{\scriptscriptstyle C})] +\exp[-\beta(A -\pi_{\scriptscriptstyle D})]}.
\label{eq_ssus_equilibrium}
\end{equation}
Under weak selection $\beta \ll 1$,
we get
\begin{equation}
x^*
\approx 
\frac{1}{2}+\frac{1}{8} \beta  (\text{$\pi_{\scriptscriptstyle C}$}-\text{$\pi_{\scriptscriptstyle D} $}),
\end{equation}
which yields
\begin{equation}
x^*
\approx 
\frac{1}{2} -\frac{1}{8} \beta  \rho
\label{eq_ssus_equilibrium_approx}
\end{equation}
which is subject to the constraint  
$0 \le x^* \le 1$. 	
The aspiration level $A$ has no impact on $x^*$
that decreases with $\beta$ and $\rho$.
The condition for cooperation to be more abundant than defection $x^* >1/2$ is
\begin{equation}
\rho <0.
\label{eq_ssus_abundance_infinite}
\end{equation}
However, the condition cannot be met
since $0 <\rho <1$ for PD games.
Thus, cooperation cannot be more abundant than defection under SSUS.
Only $x^* <1/2$ holds. 
	Although our analytical derivation of $x^* <1/2$ is based on the assumption of weak selection $\beta \ll 1$, it also holds well for strong selection $\beta \gg 1$ (Fig.\,\ref{fig_equilibrium_SSUS}). 
However, the analytical approximation of $x^*$ [Eq.\,\eqref{eq_ssus_equilibrium_approx}] works well for weak selection $\beta$,
but not so for strong selection.

\subsection{$x^*$ of SDUC}
Under the SDUC rule,
at equilibrium,
we have 
\begin{equation}
x^* 
=\frac{1+\exp[-\beta(\pi_{\scriptscriptstyle C} -A)]}{2 +\exp[-\beta(\pi_{\scriptscriptstyle C} -A)] +\exp[-\beta(A -\pi_{\scriptscriptstyle D})]}.
\label{eq_sduc_equilibrium}
\end{equation}
        
\begin{figure*}[t]         
\begin{center}     
\begin{tabular}{cc} 
 \includegraphics[width=0.43\textwidth]{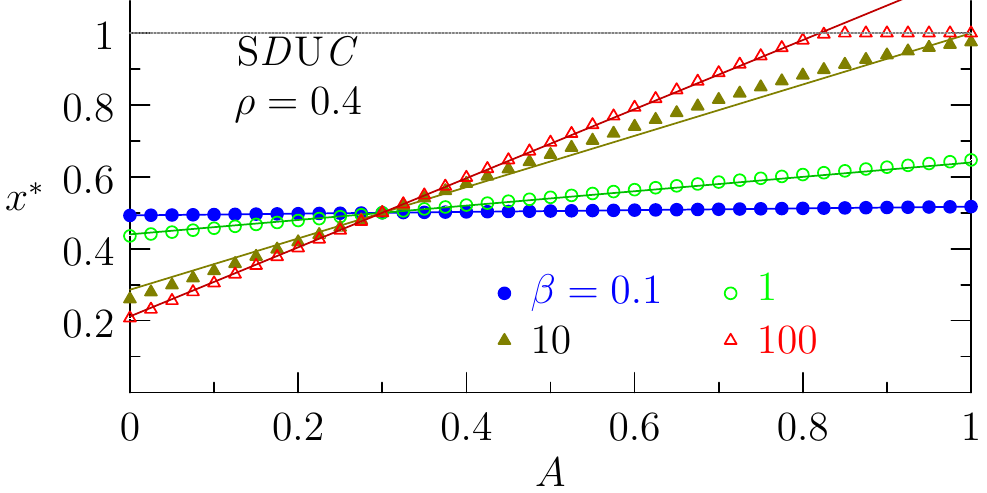}      &
\includegraphics[width=0.43\textwidth]{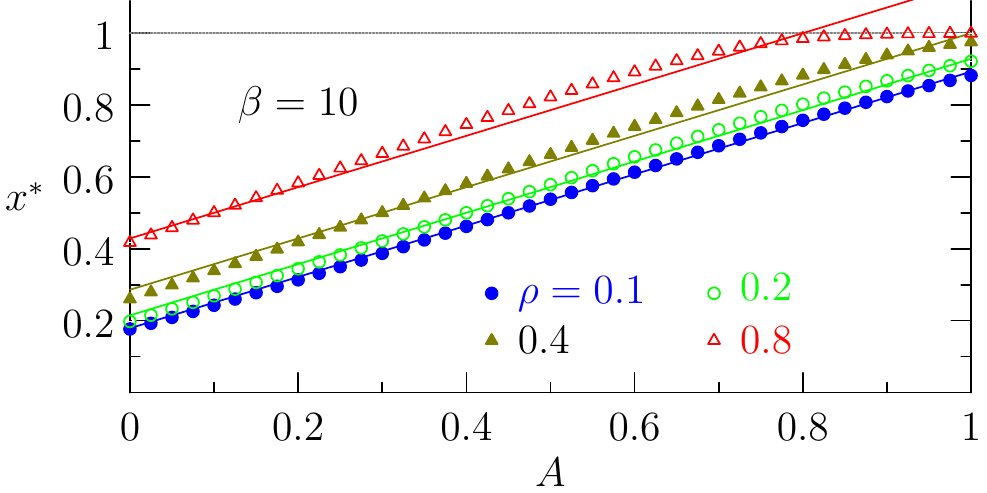}        \\
 \includegraphics[width=0.43\textwidth]{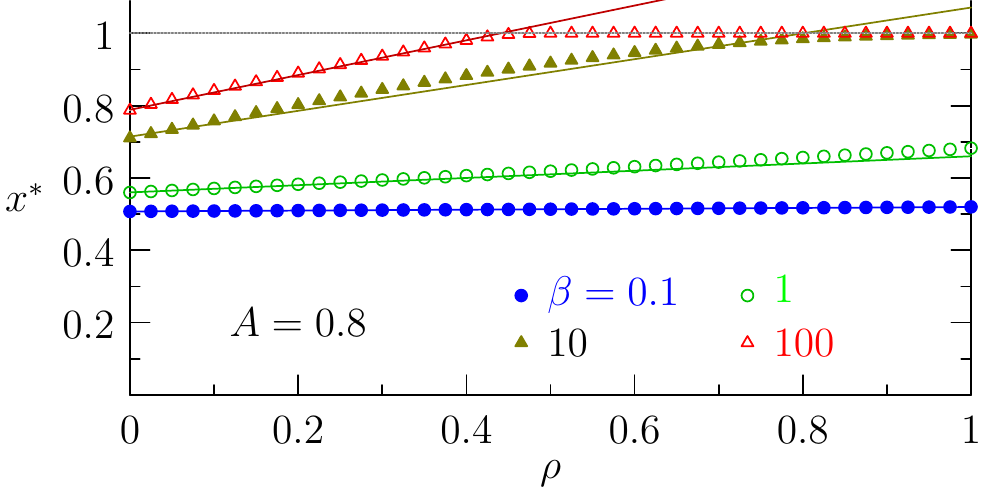}       &         
\includegraphics[width=0.43\textwidth]{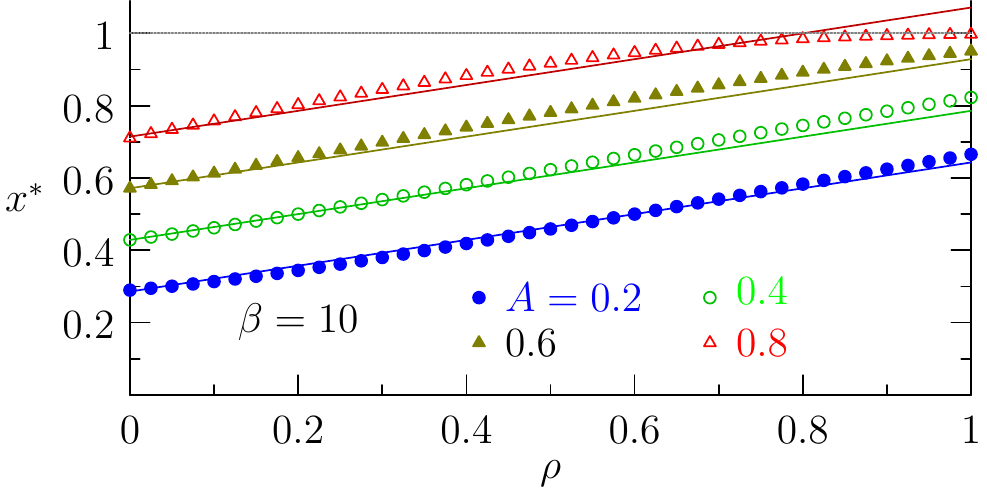}         \\      
 \includegraphics[width=0.43\textwidth]{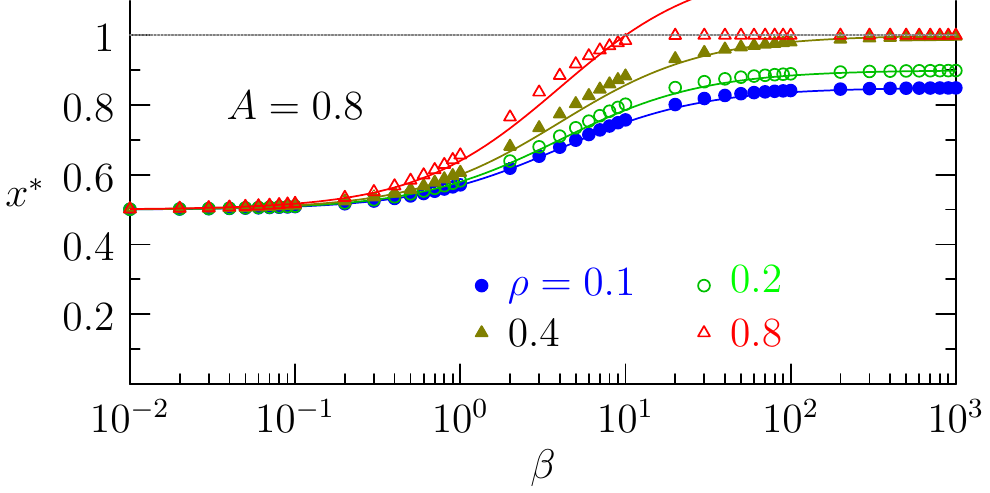}  &         
\includegraphics[width=0.43\textwidth]{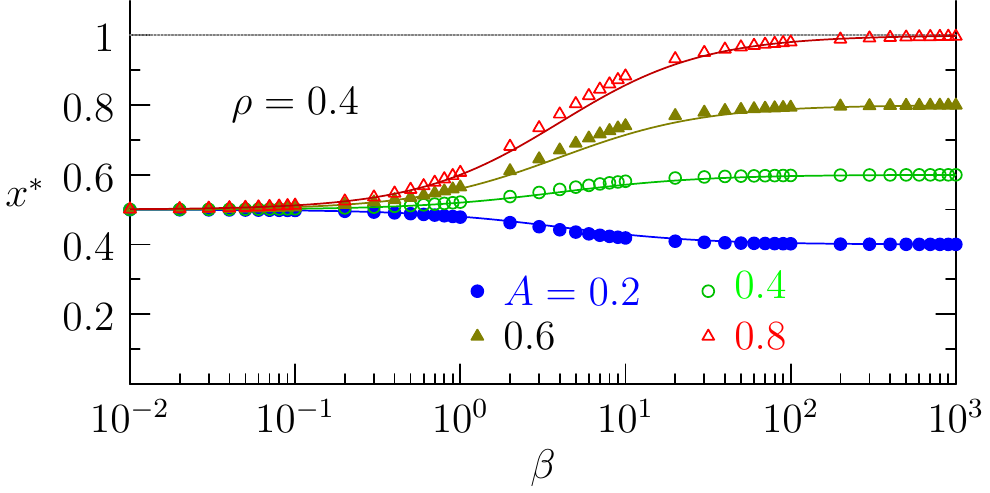}                  
\end{tabular}      
\end{center}    
\caption{ 
The equilibrium frequency $x^*$ of SDUC 
vs.\,aspiration level $A$,
cost $\rho$, and selection strength $\beta$.
The circles and triangles indicate the equilibrium obtained
by numerically solving Eq.\,\eqref{eq_sduc_equilibrium}.
The solid lines indicates the analytical approximation of the equilibrium by 
$x^* \approx 
\left[2 +(A +\rho/2)\beta\right] /\left(4 +\beta\right)
$ [Eq.\,\eqref{eq_sduc_equilibrium_approx}],
which works well even for strong selection $\beta \gg 1$.
Where $x^* \approx \left[2 +(A +\rho/2)\beta\right] /\left(4 +\beta\right) >1$,
it just needs to be capped at $x^* =1$ since $x^*$ is subject to $0 \le x^* \le1$.
}    
\label{fig_equilibrium_SDUC}   
\end{figure*}

The weak selection approximation 
	of $x^*$
is given by
\begin{equation}
x^* 
\approx 
 \frac{1}{2} -\left( \frac{\pi_{\scriptscriptstyle C} +\pi_{\scriptscriptstyle D}  -2A}{8}  \right)   \beta,
\label{eq_sduc_equilibrium_weak}
\end{equation}
which yields
\begin{equation}
x^* \approx \frac{2 +(A +\rho/2)\beta}{4 +\beta}.
\label{eq_sduc_equilibrium_approx}
\end{equation}
Note that 
the equilibrium frequency $x^*$ of SDUC [Eq.\,\eqref{eq_sduc_equilibrium_approx}] behaves in manners qualitatively different from that of SSUS [Eq.\,\eqref{eq_ssus_equilibrium_approx}].
First,
$x^*$  increases with $A$ under SDUC, whereas it does not depend on $A$ at all under SSUS.
Second,
$x^*$ increases with $\rho$ under SDUC, 
whereas it decreases under SSUS. 
Third,
as $\beta$ increases,
$x^*$ strictly increases, decreases or is constant since 	$\pdv*{x^*}{\beta} 
=4\left[A -(1 -\rho)/2\right] /(\beta +4)^2$ under SDUC,
whereas it only decreases under SSUS.
We get cooperation more abundant $x^*> 1/2$
 if the following condition is met
\begin{equation}
A >\frac{1 -\rho}{2},
\label{eq_sduc_abundance_infinite}
\end{equation} 
which is feasible.
In other words, 
cooperation can be more abundant than defection under SDUC.
The higher $A$ or $\rho$,
the easier for cooperation to be more abundant.
Although we have $x^* < 1$ for any finite $\beta$ because of the nonzero switching probabilities $q_{D\rightarrow C} >0$ and $q_{C\rightarrow D} >0$ [Eq.\,\eqref{eq_SDUC_d2c} and \ref{eq_SDUC_c2d}],
an almost full cooperation $x^* \approx 1$ is feasible
if both of the following conditions are met:
\begin{align}
A  &> \frac{2 -\rho}{2}, \\
\beta &\ge \frac{4}{(\rho +2A -2)},
\end{align}
which are derived by setting the condition for $x^* \ge 1$ from Eq.\,\eqref{eq_sduc_equilibrium_approx}. 
Although the conditions are derived under weak selection, they work well even for strong selection (Fig.\,\ref{fig_equilibrium_SDUC}). 
The higher $A$, $\rho$, or $\beta$,
the easier the almost full cooperation occurs. 
Note that 
the analytical approximation of $x^*$ [Eq.\,\eqref{eq_sduc_equilibrium_approx}] works well even for strong selection $\beta \gg 1$ under SDUC, unlike that of SUSS.

\section{Finite Populations}
The deterministic evolutionary dynamics led by aspiration assumes an infinite population.
For a finite population, we have stochastic evolutionary dynamics.
We present the stochastic dynamics of SDUC
and compare it with that of SSUS. 
The microprocess at the individual level is modeled as follows:
in each time step,
an individual is chosen at random,
who obtains its payoff in the donation game
and switches its strategy with probability $q_{\scriptscriptstyle C\rightarrow D}$ or $q_{\scriptscriptstyle D\rightarrow C}$.
For a finite population of size $N$,
the state of the population can be specified with the abundance or number of cooperators $i$.
The stochastic dynamics of the finite system can be modeled 
as a Markov chain of a one-dimensional birth-death process in discrete time.
In each time step,
the number of cooperators $i$ increases by one with probability $T^+_i$ (as a defector switches to be a cooperator),
decreases by one with probability $T^-_i$ (as a cooperator switches to be a defector),
or does not change with probability $T^0_i$.
Only these three events are possible in each time step, i.e.,\,all other transitions have zero probability. 
The transition probabilities of the Markov chain are given by
\begin{align}
T^+_i &=  \frac{N-i}{N}q_{\scriptscriptstyle D\rightarrow C}, \\
\label{eq_transition_prob_plus}
T^-_i &=  \frac{i}{N}q_{\scriptscriptstyle C\rightarrow D},  \\
T^0_i &= 1 -T^+_i -T^-_i. & &
\label{eq_transition_prob_minus}
\end{align}

\begin{figure}[t] 
\begin{center}  
\includegraphics[width=0.43\textwidth]{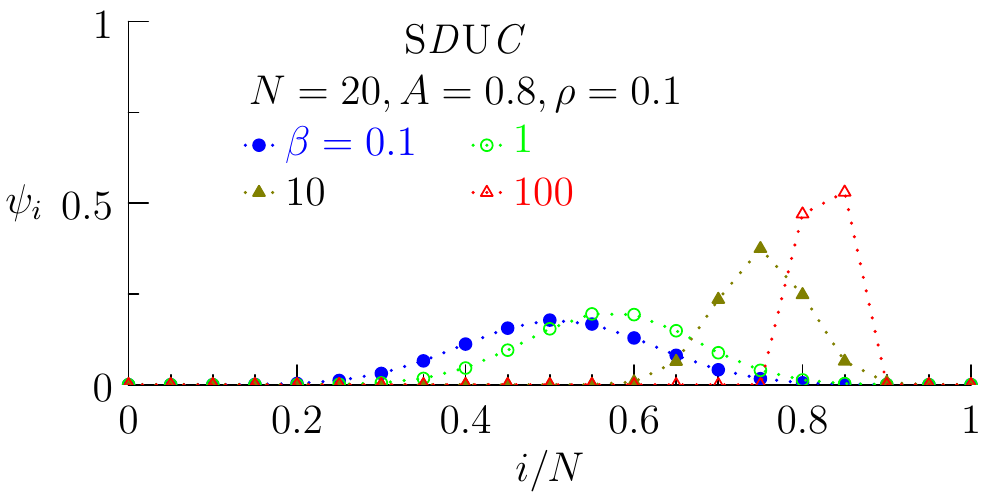}   \\               
\includegraphics[width=0.43\textwidth]{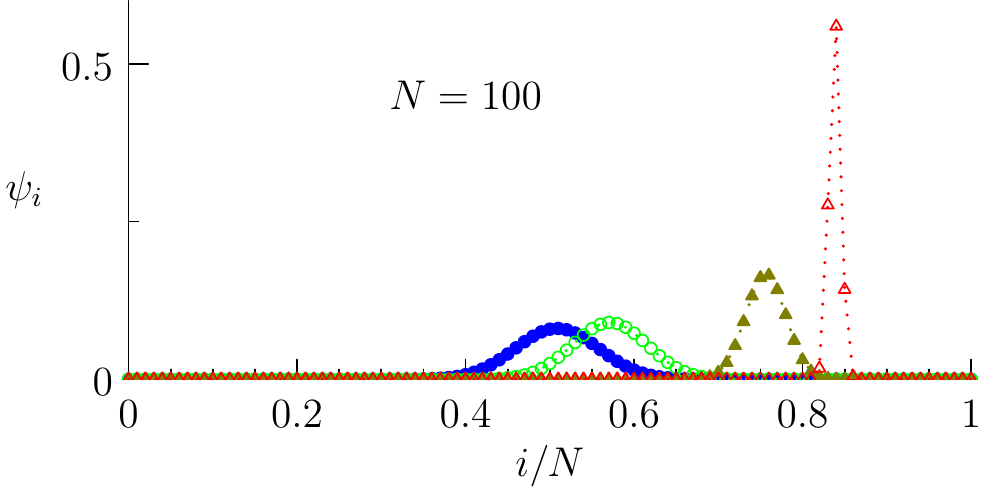} 
\end{center} 
\caption{
Stationary distributions $\psi_i$ under SDUC.
$A=0.8, \rho=0.1$.
} 
\label{fig_sduc_distribution}
\end{figure}

Let $\left(\psi_0,\ldots,\psi_j,\ldots,\psi_N\right)$ denote the stationary distribution over the abundance or number of cooperators.
In general,
the stationary distribution of a Markov chain can be obtained as the eigenvector of the transition matrix associated with the largest eigenvalue of 1.
For an one-dimensional birth-death process \cite{kampen2007stochastic, gardiner2004handbook, mahnke2009physics},
the stationary distribution is also given by 
\begin{equation}
\psi_j=
\begin{cases} 	
   \frac{1}{1+\sum_{k=1}^N \Pi^{k}_{i=1}T^{+}_{i-1} / T^{-}_i} & :  j=0\\
   \frac{\Pi^{j}_{i=1}T^{+}_{i-1} / T^{-}_i}{1+\sum_{k=1}^N \Pi^{k}_{i=1} T^{+}_{i-1} / T^{-}_i} & :  j>0.
\end{cases}
\label{eq_stat_dist}
\end{equation}
Note that for $j >0$, we have
\begin{equation}
\psi_j= \psi_0 \Pi^{j}_{i=1}T^{+}_{i-1} / T^{-}_i =\psi_{j-1} T^+_{j-1} /T^-_j
\label{eq_stat_dist_recurrence}
\end{equation}
(Fig.\,\ref{fig_sduc_distribution}).
The mean abundance of cooperation is given by
\begin{equation}
\langle X\rangle = \sum_{j=0}^N \frac{j}{N} \psi_j.
\label{eq_mean_abundance}
\end{equation}
Under weak selection $0 <\beta  \ll 1$,
 the stationary distribution $\psi_j$ can be approximated (to the first order) by
\begin{equation}
\psi_j  \approx \psi_j \big |_{\beta=0} +\pdv{\psi_j}{\beta} \bigg |_{\beta=0} \beta.
\label{eq_weak_selection}
\end{equation}

\subsection{$\langle X\rangle$ of SSUS}
Under SSUS,
a weak selection condition for cooperation to be more abundant than defection $\langle X\rangle > 1/2$ is given by
\begin{equation}
1-\rho + (-\rho) > 1 + 0,
\label{eq_ssus_abundance}
\end{equation}
i.e.,\,the sum of payoff entries for cooperation should be larger than that for defection.
For the derivation of Eq.\,\eqref{eq_ssus_abundance},
see Ref.\,\cite{du2015aspiration} 
where the condition for $\langle X\rangle > 1/2$ was analytically derived,
but not $\langle X\rangle$ itself.
However,
the condition of Eq.\,\eqref{eq_ssus_abundance} is equivalent to $\rho <0$,
that is the same as that of infinite populations [Eq.\,\eqref{eq_ssus_abundance_infinite}]
and cannot be met.
Under SSUS,
cooperation cannot be more abundant than defection in finite populations nor infinite populations.

\subsection{$\langle X\rangle$ of SDUC}
\begin{figure}[t] 
\begin{center}
\includegraphics[width=0.43\textwidth]{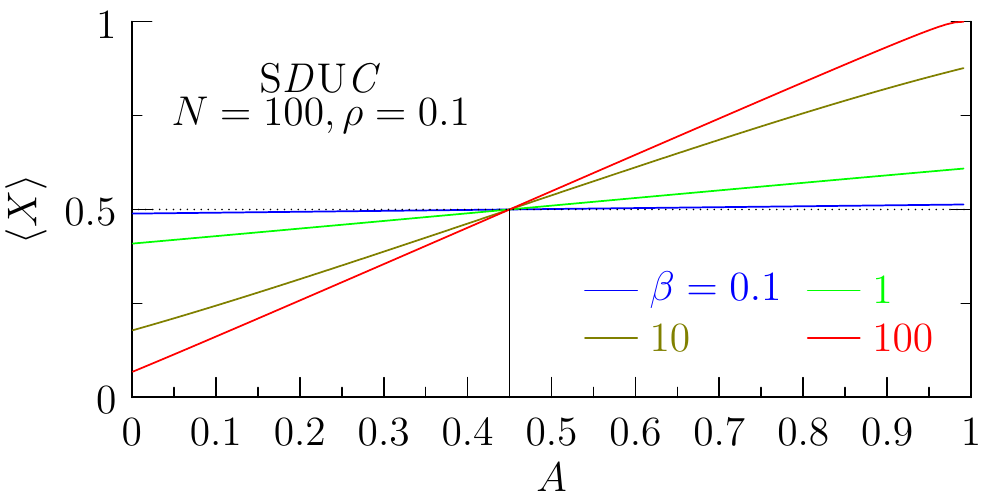}
\end{center} 
\caption{
Mean abundance $\langle X \rangle$ vs.\,aspiration level $A$.
$\langle X \rangle$ is computed numerically.
The horizontal (dotted) line corresponds to $\langle X \rangle =1/2$
and the vertical line, $A  =(1 -\rho)/2 =0.45$ where $\rho=0.1$.
It clearly demonstrates the validity of Eq.\,\eqref{eq_sduc_abundance} as the condition for $\langle X\rangle >1/2$,
which works well even for strong selection $\beta =100$.
\label{fig_sduc_abundance}
} 
\end{figure}

Under SDUC, the transition probabilities are given by
\begin{align}
T^+_i &= \frac{N-i}{N} \frac{1}{1 +e^{-\beta\left[ A  -\pi_D(i)\right]}},
\label{eq_tplus_SDUC}\\
T^-_i &= \frac{i}{N} \frac{1}{1 +e^{-\beta\left[\pi_C(i) - A \right]}}.
\label{eq_tminus_SDUC}
\end{align}
Under weak selection,
we get an analytical approximation of the mean abundance by
\begin{equation}
\langle X\rangle 
\approx \frac{1}{2} +\frac{1}{8} \left(2A +\rho -1\right) \beta. 
\label{eq_mean_abudance_SDUC}
\end{equation}
For the derivation of Eq.\,\eqref{eq_mean_abudance_SDUC}, see Appendix \ref{deriv_mean_abudance_SDUC}.
The condition for cooperation to be more abundant than defection $\langle X\rangle > 1/2$ is then given by
\begin{equation}
A  > \frac{1 -\rho}{2}.
\label{eq_sduc_abundance}
\end{equation} 
Although the condition [Eq.\,\eqref{eq_sduc_abundance}] is derived under weak selection, it works well for strong selection (Fig.\,\ref{fig_sduc_abundance}).
Note that the condition of finite populations [Eq.\,\eqref{eq_sduc_abundance}] is the same as that of infinite populations [Eq.\,\eqref{eq_sduc_abundance_infinite}].
Under SDUC,
cooperation can be more abundant than defection
 in both finite and infinite populations under the same condition.

\subsection{Correspondence between stochastic and deterministic dynamics}
For infinite populations,
the abundance of cooperation is captured by the equilibrium frequency $x^*$
that can be analytically approximated
and we straightforwardly derive the condition for $x^* >1/2$  from it.
For finite populations,
the abundance of cooperation is captured by the mean $\langle X \rangle$ of the stationary distribution.
Although the condition for $\langle X \rangle >1/2$ was analytically derived,
$\langle X \rangle$ itself was not so
in the previous work 
\cite{du2015aspiration, du2014aspiration}.
The lack of an analytical representation of $\langle X \rangle$ limits further understanding of the aspiration dynamics of finite populations.
It also 	
makes it difficult to link the stochastic dynamics of a finite population and the deterministic dynamics of an infinite population \cite{du2014aspiration}. 
One could consider the latter as a limit case of the former
as the population size increases to infinity

\begin{figure*}[t]  
\begin{center}     
\begin{tabular}{cc} 
 \includegraphics[width=0.43\textwidth]{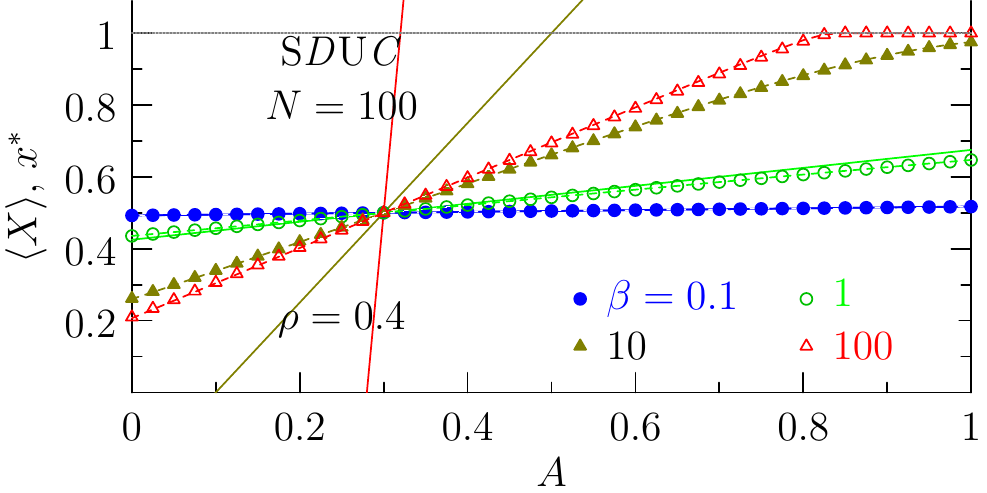}       &
\includegraphics[width=0.43\textwidth]{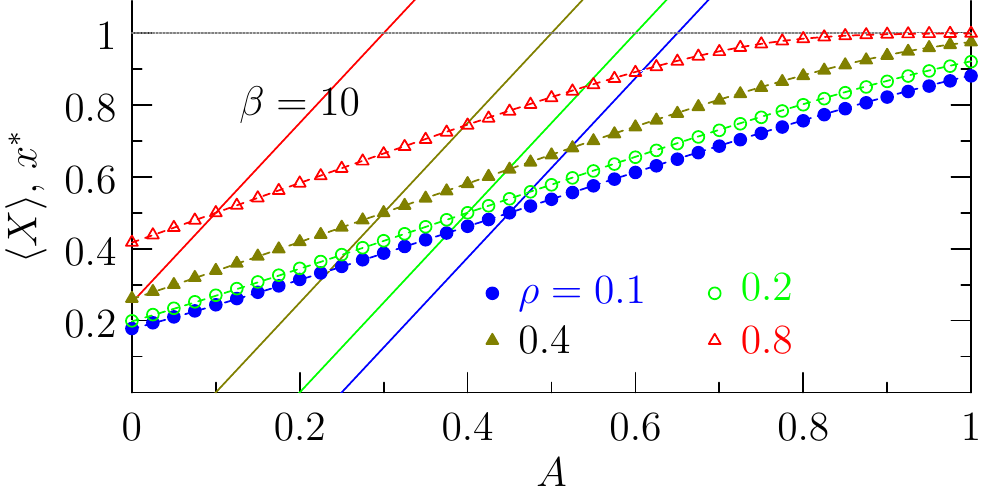}       \\   
 \includegraphics[width=0.43\textwidth]{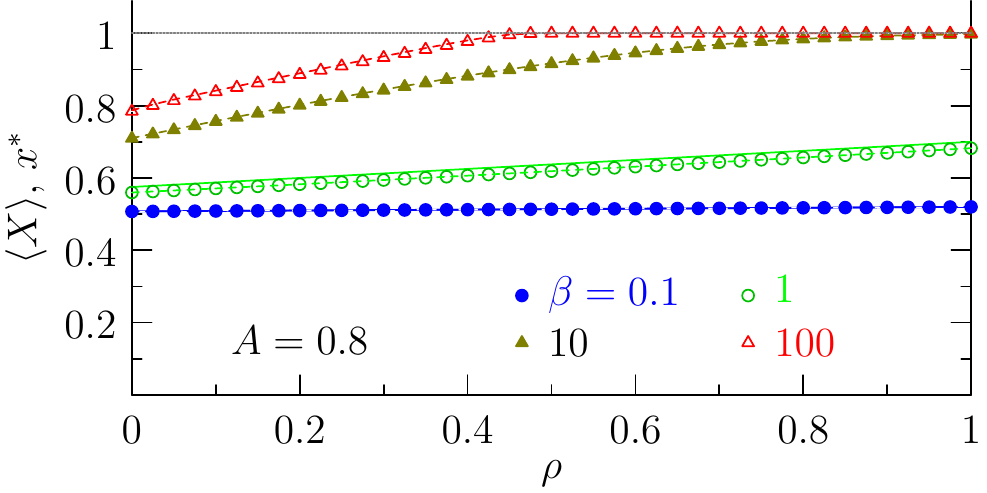}  &     
\includegraphics[width=0.43\textwidth]{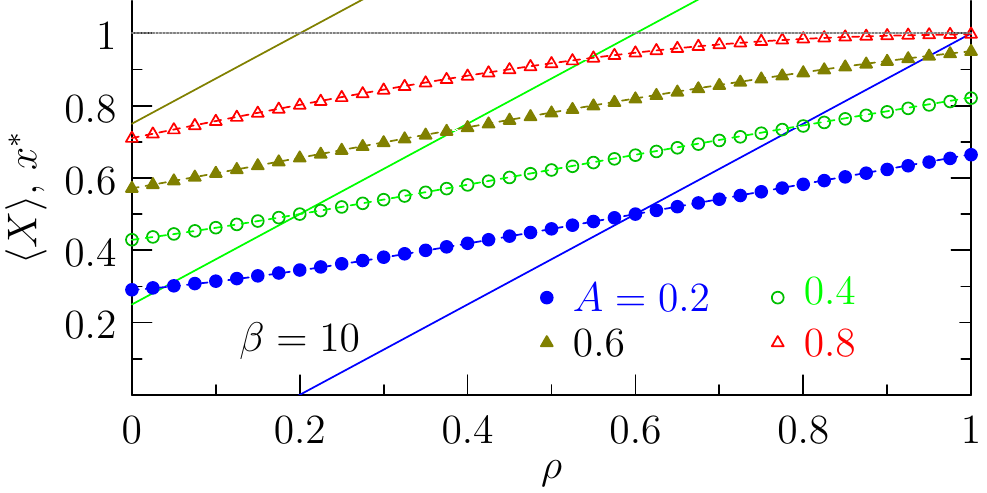}       \\                           
 \includegraphics[width=0.43\textwidth]{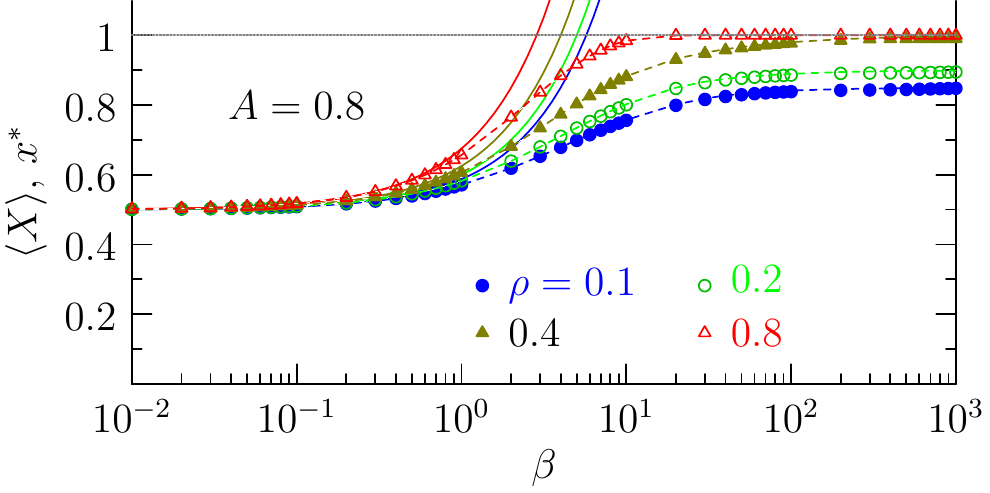}   &     
\includegraphics[width=0.43\textwidth]{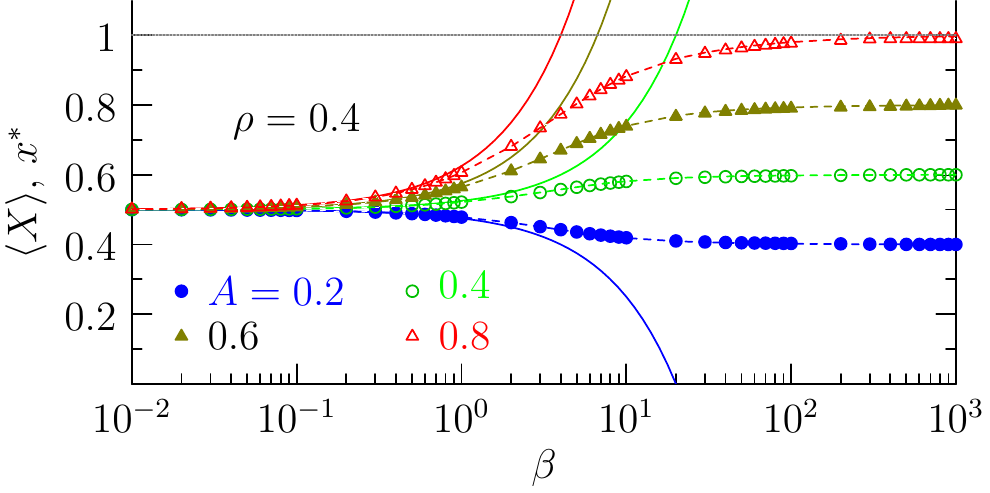}          
\end{tabular}    
\end{center}  
\caption{
Mean abundance $\langle X \rangle$ of SDUC vs.\,aspiration level $A$,
cost $\rho$, and selection strength $\beta$.
The population size is $N=100$.
The circles and triangles indicate $\langle X \rangle$ numerically obtained.
The dashed curves represents the equilibrium frequency $x^*$ in an infinite population,
which is obtained by numerically solving Eq.\,\eqref{eq_sduc_equilibrium}
as in Fig.\,\ref{fig_equilibrium_SDUC}.
$\langle X\rangle$ and $x^*$ numerically well match each other.
The solid curves represent the analytical approximations by Eq.\,\eqref{eq_approx_linear},
which work well only for weak selection $\beta$. 
} 
\label{fig_mean_SDUC}
\end{figure*}

In our work,
we analytically approximate $\langle X \rangle$ 
[Eq.\,\eqref{eq_mean_abudance_SDUC}],
that not only yields the condition for $\langle X \rangle >1/2$ [Eq.\,\eqref{eq_sduc_abundance}] in a straightforward manner, 
but also provides further insights on the stochastic dynamics of SDUC in finite populations.
According to Eq.\,\eqref{eq_mean_abudance_SDUC},
for instance,
$\langle X \rangle \approx1/2 +\left(2A +\rho -1\right)\beta/8$ increases with $A$ and $\rho$.
According to Eq.\,\eqref{eq_sduc_equilibrium_approx},
this is qualitatively similar to $x^* \approx  \left[4 +(2A +\rho)\beta\right]/\left[2(4 +\beta)\right]$ 
in that the latter also increases with $A$ and $\rho$ in infinite populations.
However,
the analytical approximations Eqs.\,\eqref{eq_mean_abudance_SDUC} and \eqref{eq_sduc_equilibrium_approx} of $\langle X \rangle$ and $x^*$
do not match each other
whereas numerically computed $\langle X \rangle$ and $x^*$ do so.
One way to resolve this incompatibility between the analytical approximations would be to linearize Eq.\,\eqref{eq_sduc_equilibrium_approx} in $\beta$ by
\begin{equation}
x^* \approx \frac{1}{2} +\frac{1}{8}\left(2A +\rho -1\right)\beta.
\label{eq_equi_linearization}
\end{equation}
Then we get
\begin{equation}
\langle X \rangle \approx x^* \approx \frac{1}{2} +\frac{1}{8}\left(2A +\rho -1\right)\beta
\label{eq_approx_linear}
\end{equation}
and we analytically establish a quantitative correspondence between the dynamics of finite and infinite populations.
However, the correspondence is achieved at a cost of approximation accuracy of Eq.\,\eqref{eq_approx_linear},
which works well only for weak selection (Fig.\,\ref{fig_mean_SDUC}).

While keeping Eq.\,\eqref{eq_sduc_equilibrium_approx} that well approximates $x^*$ even for strong selection (Fig.\,\ref{fig_equilibrium_SDUC}),
we provide
a better alternative 
in analytically establishing the quantitative correspondence between the dynamics of finite and infinite populations.
Rather than the mean $\langle X \rangle$,
we capture the abundance of cooperation by the mode of the stationary distribution, 
i.e.,\,the number (of cooperators) that occurs most frequently where the distribution peaks.
Our approximation $i^*$ of the mode 
satisfies
\begin{equation}
T^+_{i^*-1} /T^-_{i^*} =1,
\label{eq_sduc_mode_1}
\end{equation}
which yields
\begin{equation}
\frac{i^*}{N +1}  =\frac{1 +e^{-\beta\left[\pi_C(i^*) - A \right]}}{2 +e^{-\beta\left[\pi_C(i^*) - A \right]} +e^{-\beta\left[ A  -\pi_D(i^*-1)\right]}}
\label{eq_sduc_mode}
\end{equation}
For the derivation of Eq.\,\eqref{eq_sduc_mode_1} and \ref{eq_sduc_mode},
see Appendix \ref{deriv_sduc_mode}.
Equation \eqref{eq_sduc_mode} is a finite analog of Eq.\,\eqref{eq_sduc_equilibrium}.    
Note that $i^*$ is a real number approximation of the (integer) mode $\lfloor i^* \rfloor$, 
the most frequent number of cooperators in a finite population
where $\lfloor i^* \rfloor$ denotes the largest integer that is less than or equal to $i^*$.
In slight abuse of notation, we will write $i^*$ in place of $\lfloor i^* \rfloor$.
What we are interested in is 
the (normalized) mode $\lfloor i^* \rfloor /N$ 
that is well approximated by $i^* /N$ 
since $i^*/N -\lfloor i^* \rfloor /N <1/N$ is negligible for a large $N$. 
From Eq.\,\eqref{eq_sduc_mode}
under weak selection $\beta \ll 1$, 
we get
\begin{equation}
\frac{i^*}{N +1} \approx \frac{1}{2} -\frac{1}{8}\left[\pi_C(i^*) +\pi_D(i^*-1) -2A\right] \beta,
\label{eq_sduc_mode_weak_selection}
\end{equation}
which yields
\begin{equation}
\frac{i^*}{N+1} \approx\frac{N-1}{4(N-1) +(N+1)\beta} \left[2 +\left(A  +\frac{\rho}{2} +\frac{1}{N-1}\right) \beta\right].
\label{eq_sduc_mode_approx}
\end{equation}
 
\begin{figure}[t]    
\begin{center}
\begin{tabular}{c} 
\includegraphics[width=0.43\textwidth]{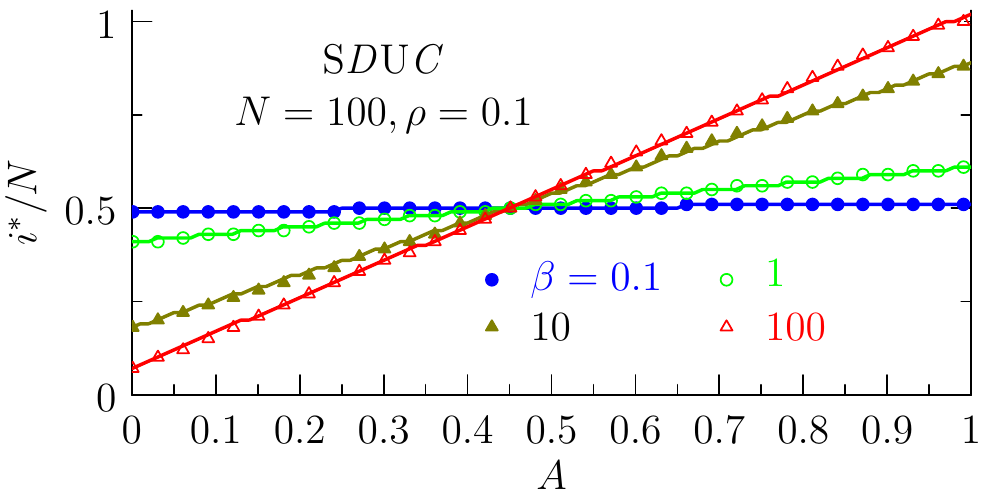}            
\end{tabular}
\end{center} 
\caption{   
Mode $i^* /N$ of SDUC vs.\,aspiration level $A$.
The circles and triangles denote 
the mode numerically obtained by locating the peak of the discrete distribution $\psi_i$.
The solid curves represent the analytical approximation of the mode by Eq.\,\eqref{eq_sduc_mode_approx}.
The analytical approximation well fits the ground truth even for strong selection $\beta=100$. 
}  
\label{fig_sduc_distribution_mode}
\end{figure}

\begin{figure*}[t] 
\begin{center}      
\begin{tabular}{cc} 
 \includegraphics[width=0.43\textwidth]{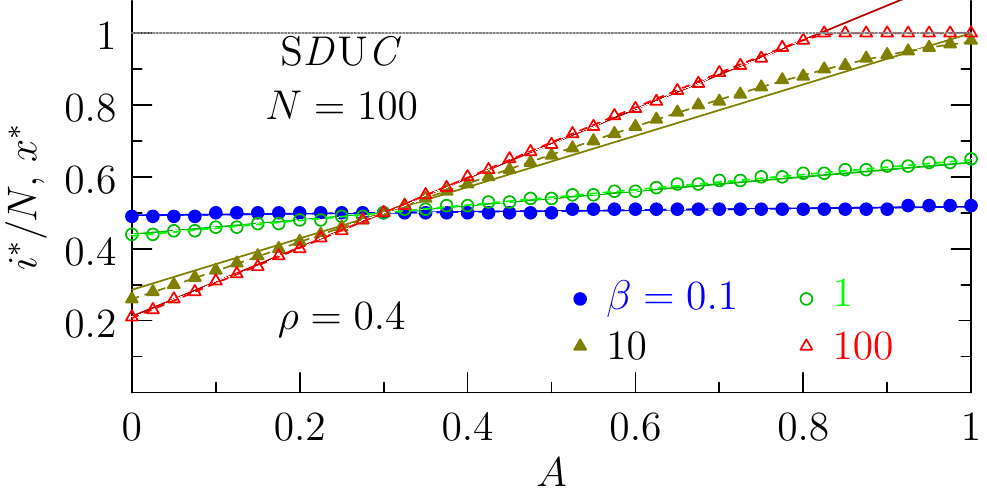}       &
\includegraphics[width=0.43\textwidth]{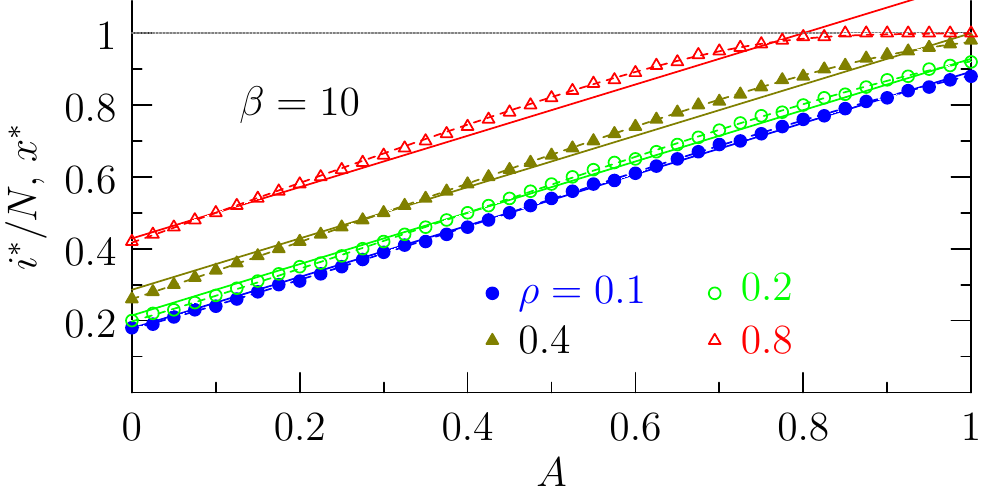}       \\   
 \includegraphics[width=0.43\textwidth]{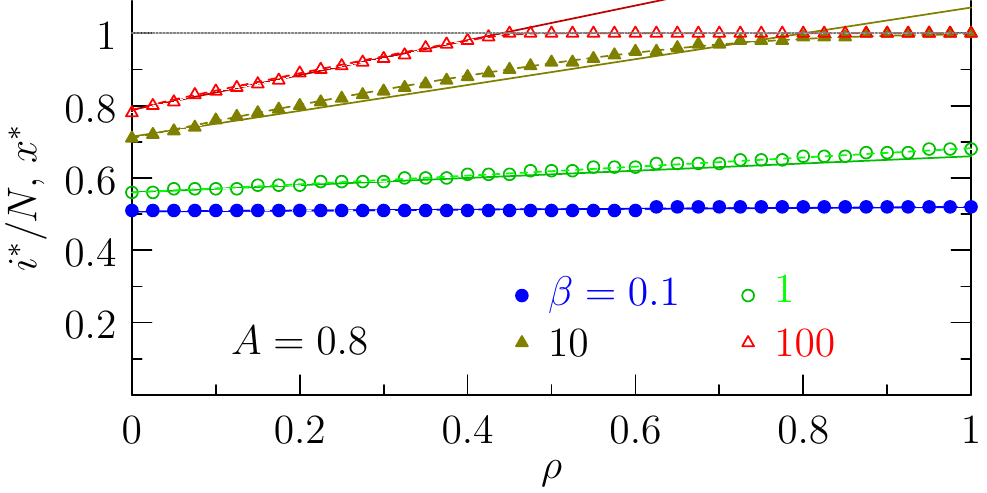}  &     
\includegraphics[width=0.43\textwidth]{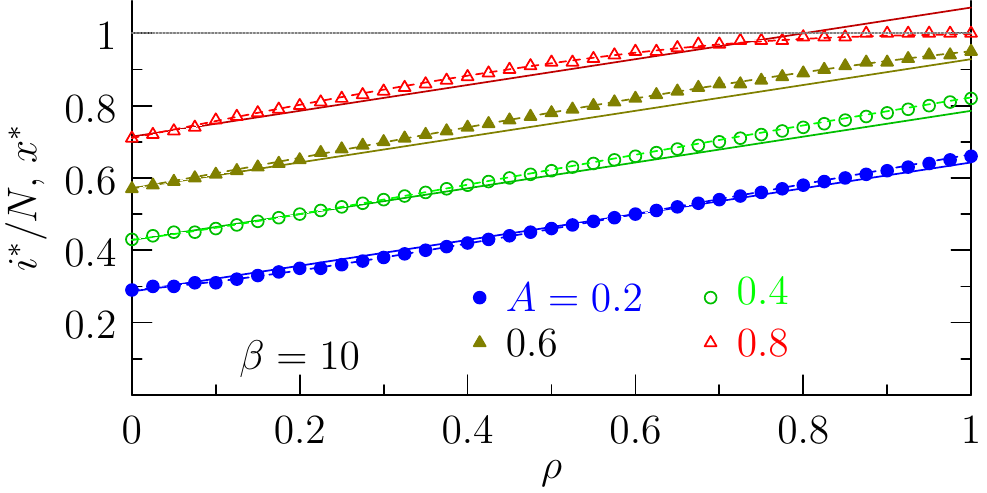}       \\                           
 \includegraphics[width=0.43\textwidth]{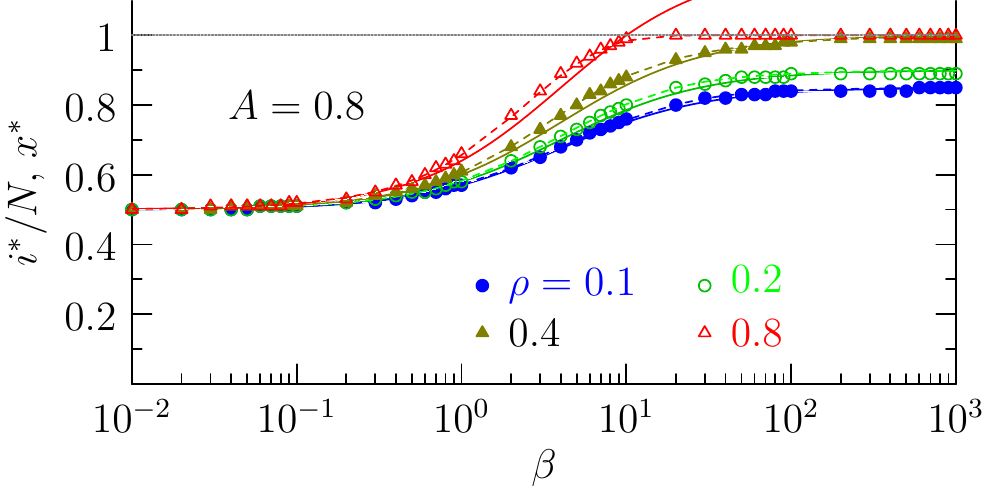}   &     
\includegraphics[width=0.43\textwidth]{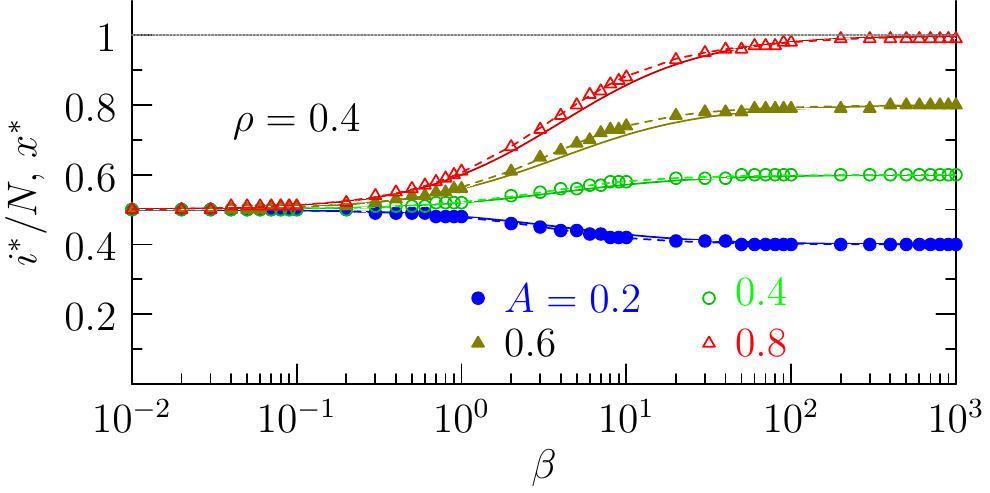}          
\end{tabular}    
\end{center} 
\caption{
Mode $i^*/N$ of SDUC vs.\,aspiration level $A$, cost $\rho$ and selection strength $\beta$.
The population size is $N=100$.
The circles and triangles indicate $i^*/N$ numerically obtained.
The dashed curves represent the equilibrium frequency $x^*$ in an infinite population,
which is obtained by numerically solving Eq.\,\eqref{eq_sduc_equilibrium}
as in Fig.\,\ref{fig_equilibrium_SDUC}.
$i^*/N$ and $x^*$ well matches each other.
The solid curves represent the analytical approximations by Eq.\,\eqref{eq_approx_rational},
which work well even for strong selection $\beta$.
} 
\label{fig_mode_SDUC}
\end{figure*}

Note that Eqs.\,\eqref{eq_sduc_mode_weak_selection} and \eqref{eq_sduc_mode_approx} are finite analogs of Eqs.\,\eqref{eq_sduc_equilibrium_weak} and \eqref{eq_sduc_equilibrium_approx}, respectively (Fig.\,\ref{fig_sduc_distribution_mode}).
For a large population size $N$, 
Eq.\,\eqref{eq_sduc_mode_approx} is simplified by
\begin{equation}
\frac{i^*}{N} \approx \frac{2 +(A +\rho/2)\beta}{4 +\beta}
\label{eq_sduc_mode_approx_large}.
\end{equation}
From Eqs.\,\eqref{eq_sduc_mode_approx_large} and \eqref{eq_sduc_equilibrium_approx},
we have
\begin{equation}
\frac{i^*}{N} \approx x^* \approx \frac{2 +(A +\rho/2)\beta}{4 +\beta}.
\label{eq_approx_rational}
\end{equation}
Using the analytical approximation of a high accuracy,
we are able to link the dynamics of finite populations
to that of infinite populations 
where the mode (of a stationary distribution) of the stochastic dynamics corresponds to the equilibrium frequency of the deterministic dynamics (Fig.\,\ref{fig_mode_SDUC}).

\subsection{Analytical approximations of stationary distributions}
  
\begin{figure}[t]  
\begin{center}
\includegraphics[width=0.43\textwidth]{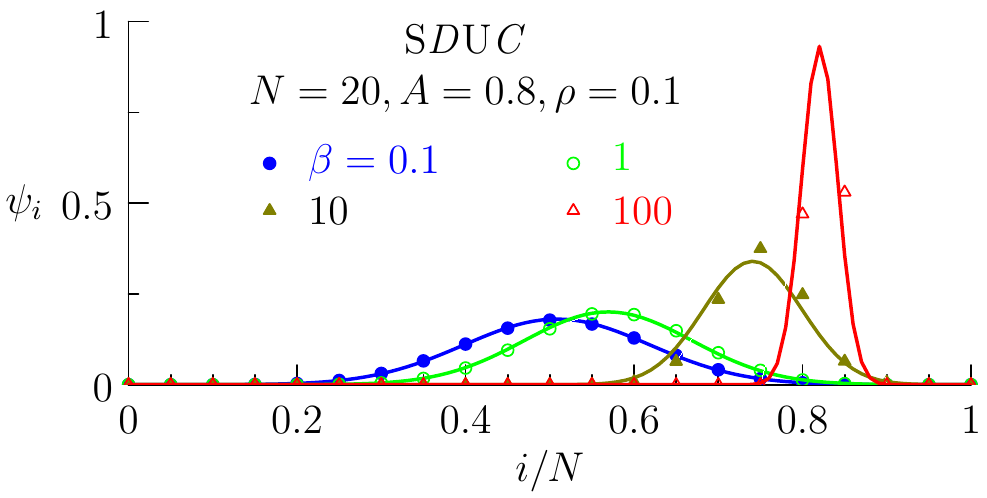}        \\ 
\includegraphics[width=0.43\textwidth]{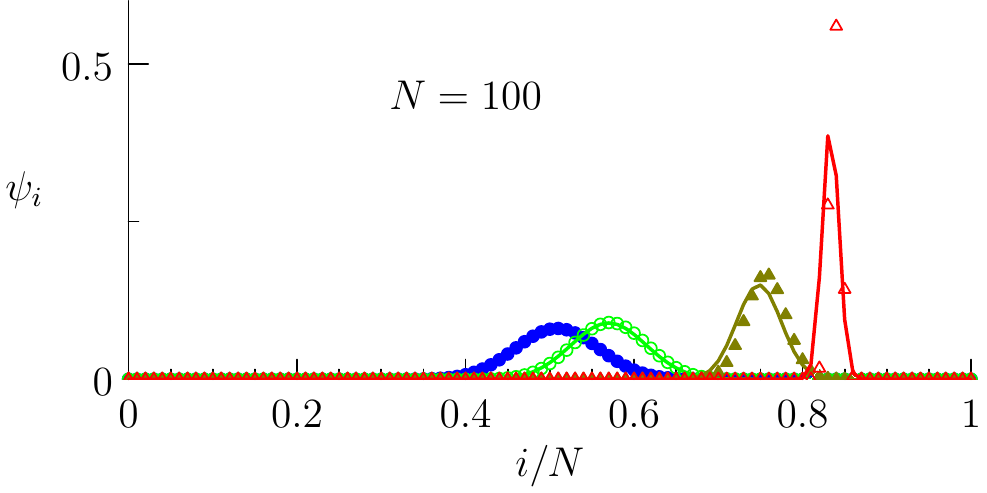}           
\end{center}  
\caption{
The stationary distributions $\psi_i$ are well  approximated by the normal distributions $\mathcal{N}_{\mu,\sigma}(i)$.
The circles and triangles indicate $\psi_i$ that is numerically obtained.
The solid curves indicate the normal distributions $\mathcal{N}_{\mu,\sigma}(i)$.
}  
\label{fig_sduc_distribution_approx}
\end{figure}

\begin{figure}[t] 
\begin{center}
\begin{tabular}{c}
\includegraphics[width=0.43\textwidth]{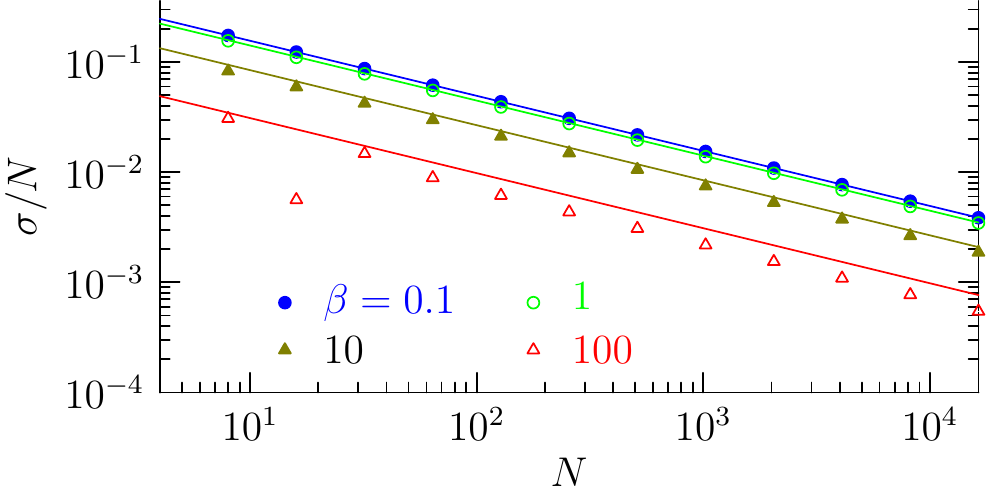}         
\end{tabular}
\end{center}  
\caption{
Normalized standard deviation $\sigma/N$ vs.\,population size $N$.
The solid curves denote analytical approximation of the deviation [Eq.\,\eqref{eq_deviation}].
The circles and the triangles denote the deviation numerically computed.
The deviation decreases as the population size increases.
}  
\label{fig_sduc_deivation_approx}
\end{figure}

\begin{figure}[t] 
\begin{center}
\includegraphics[width=0.43\textwidth]{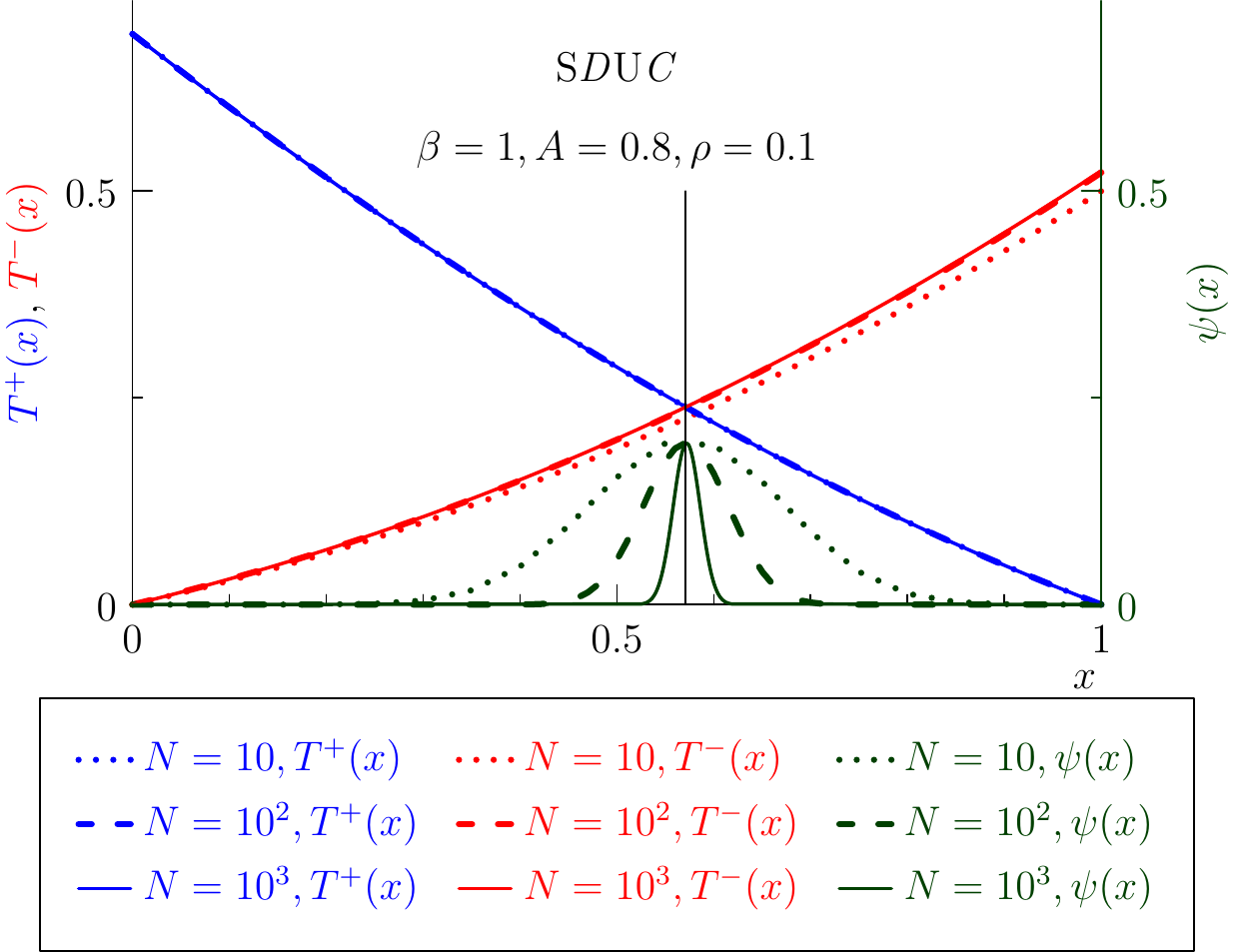}        
\end{center}  
\caption{
The transition probabilities and the stationary distribution over $x=i/N$
where  $T^+(x) = T^+_i, T^-(x) =T^-_i$ and $\psi(x) = \psi_i$.
The vertical line indicates the location of the equilibrium $x^*$ of the deterministic dynamics, 
which well matches the mode $i^*/N$ of the stationary distributions as well as $x=x^*$ such that $T^+(x^*) =T^-(x^*)$.
For visual clarity, the graphs of distributions $\psi(x)$ are scaled such that each of them has the same height at the mode.
} 
\label{fig_correspondence}
\end{figure}

The mean and the  mode would have a less predictive meaning
if the deviation of the stationary distribution is relatively large \cite{traulsen2007pairwise}. 
To estimate the deviation,
under weak selection,
we analytically approximate the stationary distribution
by
\begin{equation}
	\psi_i / \psi_0 
	\propto
	 \exp[-\frac{(i -\mu)^2}{2\sigma^2}]
\label{eq_dist_approx} 
\end{equation}
where 
\begin{align}
\mu &=\frac{N(N-1)}{4(N-1) +N\beta}\left(2 +\left[A +\frac{\rho}{2}+\frac{1}{2(N-1)}\right]\beta\right),
\label{eq_mean}\\
\sigma^2   &=\frac{N(N-1)}{4(N-1) +N\beta}.
\label{eq_deviation}
\end{align}
For the derivation of Eq.\,\eqref{eq_dist_approx}, see Appendix \ref{deriv_dist_approx}. 
The stationary distribution $\psi_i$ is thus
approximated
by a normal distribution $\mathcal{N}_{\mu,\sigma}(i)$
of mean $\mu$ and standard deviation $\sigma$.
Note that 
the mean $\mu$ of the normal distribution well approximates the mode $i^*$ of the stationary distribution [Eq.\,\eqref{eq_sduc_mode_approx}] (Fig.\,\ref{fig_sduc_distribution_approx}).
For a large $N$,
Eqs.\,\eqref{eq_mean} and \eqref{eq_deviation}
are simplified by
\begin{align}
\frac{\mu}{N} & \approx \frac{2 +(A +\rho/2 \beta}{4 +\beta},
\label{eq_mean_approx} \\
\frac{\sigma}{N} &\approx \frac{1}{\sqrt{N\left(4 +\beta\right)}}.
\label{eq_sduc_std_deviation}
\end{align}
Eq.\,\eqref{eq_mean_approx} of $\mu/N$ well matches Eq.\,\eqref{eq_approx_rational} of $i^*/N$ and $x^*$.
Because the standard deviation is relatively small  for a large $N$,
the mean and the mode of the distribution have a predictive meaning. 
As the population size $N \rightarrow \infty$, especially,
the (normalized) standard deviation $\sigma /N$ vanishes with $1/\sqrt{N}$
and the distribution thus converges to a delta function that peaks at $x^*$
where stochastic fluctuations are suppressed (Figs.\,\ref{fig_sduc_deivation_approx} and \ref{fig_correspondence}). 
In contrast to the previous work \cite{du2014aspiration, du2015aspiration},
we analytically show the link 
between the stochastic aspiration dynamics of a finite population and the deterministic dynamics of an infinite population,
the latter of which is taken as a limit case of the former 
as the population size increases to  infinity.

\section{Summary and Discussion}

Imitation-based strategy updates  have been widely used to study the evolution of cooperation in social dilemmas.
However, the results of the recent behavioral experiments question the applicability of the imitation dynamics 
and it is thus well worth considering alternatives such as aspiration dynamics.
In PD games,
conventional aspiration dynamics yields the coexistence of cooperation and defection at equilibrium in an infinite population
and a nontrivial stationary distribution in a finite population.
In contrast, imitation dynamics yields the extinction of cooperation in a well-mixed infinite population
and fixation in a finite population.

The details of the update rules can have significant impacts on the evolutionary outcomes
and numerous variations in imitation dynamics have been studied such as
birth-death, death-birth, Moran process, pairwise comparison, imitate-the-best, etc.
To our knowledge,
however,  few variations exist in aspiration dynamics.  
There are additional reasons to seek	 alternative mechanisms of strategy updates in aspiration dynamics.
Although aspiration-led strategy updates are often observed in studies of both animal and human behavioral ecology,
not all of them comply with the conventional aspiration-led update by SSUS.
Being conditioned on the current strategy 
as well as the payoff-aspiration difference,
SSUS yields a different strategy depending on the current strategy.
When the aspiration level is met,
for instance, 
some individuals (continue to) cooperate while others defect according to SSUS.
In animal behavior,
however,
a strategy update is often conditioned on only the payoff-aspiration difference, 
but not the current strategy.
In the variance-sensitive foraging behavior of animals, 
for example, 
whether animals choose a variance-averse strategy or a variance-prone strategy is entirely conditioned on whether an aspiration level is met or not, but not on the current strategy \cite{Caraco:1980aa, lim2015origin}.
SSUS is a kind of reinforcement learning
that assumes 
humans to do less or abandon the strategy diminishing in value
and switch to the other strategy potentially more rewarding.
However, humans sometimes show an opposing tendency,
trying harder at what they have been doing rather than less \cite{Rabbitt:1966aa, Laming:1979aa, Gratton:1992aa}.
In the context of aspiration-led strategy updates,
this implies that individuals do not necessarily switch the current strategy
even if the aspiration is not met, contrary to SSUS.

With these motivations behind,
we search the whole space of strategy update rules led by aspiration
to derive a rule
that meets the desirable properties.
Previously,
a space of conditional cooperative strategies 
was searched to derive desirable strategies in imitation dynamics \cite{Hilbe:2017fk}.
Rather than a space of 
 strategies in imitation dynamics,
we search a space of 
strategy update mechanisms in aspiration dynamics
and introduce SDUC as an alternative
to SSUS.
Depending on the 
payoff-aspiration difference,
SDUC specifies which strategy to ``take'',
while SSUS specifies whether to ``switch'' the current strategy.
SDUC seems psychologically intuitive
in that 
 individuals opt for the costly pro-social action of cooperation only when they are in need
and opt against it otherwise.

For an infinite population,
we get an analytical approximation of
the abundance of cooperation at equilibrium
for SSUS and SDUC, respectively.
From the equilibrium abundance,
we can straightforwardly derive the condition that yields more abundant cooperation  than defection.
SDUC is simpler than SSUS
in that the strategy update of SDUC is conditioned on the payoff-aspiration difference
but not on the current strategy,
whereas SSUS is conditioned on both.
However, SDUC yields more variety in the evolutionary outcomes of PD games
than SSUC does, the latter of which only yields cooperation to be less abundant.
SDUC can yield cooperation more abundant than defection and vice versa.
SDUC can even lead to  almost full cooperation.      
        
For a finite population,
the previous works analytically derived the condition for  more abundant cooperation,
but not the abundance of cooperation itself \cite{du2015aspiration, du2014aspiration}.
The lack of the analytical representation of the abundance limits further understanding of  aspiration dynamics of finite populations
and causes  difficulty in linking the dynamics between finite and infinite populations.
In our work,
we derive the analytical representations of the abundance of cooperation as well as the stationary distributions in finite populations for SDUS.
From the analytical representations,
we straightforwardly derive the condition for more abundant cooperation 
and link the stochastic dynamics of finite populations
to the deterministic dynamics of infinite populations,
the latter of which is considered as a limit case of the former as the population size increases to infinity.

SSUS and SDUC also yield differences in terms of the relation between 
cooperation and cost.
The abundance of cooperation under SSUS decreases with the cost of cooperation,
which also corroborates the outcome of imitation-led evolutionary dynamics \cite{Ohtsuki:2006fk, Baalen:1998ve}.
On the other hand,
the abundance of cooperation under SDUC increases with the cost,
which appears somewhat counter-intuitive.
When more realistic ecological factors are taken into consideration, 
however,
similar positive correlations between cooperation and cost occur due to the spatial self-organization in imitation-led eco-evolutionary dynamics \cite{szolnoki2014binary, Colizzi:2016ty, smaldino2013increased}.
While the eco-evolutionary dynamics requires additional complexities such as a structured population, a nonconstant population size and movements of individuals,
SDUC yields the positive correlation in a minimal model
that  assumes only a well-mixed population of a constant size and does not require movements of individuals.

We hope that the introduction of SDUC paves a way of searching for further mechanisms of aspiration-led strategy updates.
For instance, the space of possible update rules could be expanded
by taking the cost of cooperation into account in addition to the aspiration level.

\appendix

\section{Derivation of Eq.\,\eqref{eq_mean_abudance_SDUC}
\label{deriv_mean_abudance_SDUC}}

We seek the condition that yields cooperation to be more abundant than defection $\langle X \rangle > 1/2$ under weak selection $0 < \beta \ll 1$.
According to Eqs.\,\eqref{eq_mean_abundance} and \eqref{eq_weak_selection},
the mean abundance $\langle X \rangle$  can be approximated by
\begin{equation}
\langle X\rangle = \sum_{j=0}^N \frac{j}{N} \psi_j
\approx \sum_{j=0}^N \frac{j}{N} \left(\psi_j \big |_{\beta=0} +\pdv{\psi_j}{\beta} \bigg |_{\beta=0} \beta\right).
\label{eq_mean_abundance_approx}
\end{equation}

The stationary distribution $\psi_j$ is given by Eq.\,\eqref{eq_stat_dist}:
\[
\psi_j=
\begin{cases} 	
   \frac{1}{1+\sum_{k=1}^N \Pi^{k}_{i=1}T^{+}_{i-1} / T^{-}_i} & :  j=0\\
   \frac{\Pi^{j}_{i=1}T^{+}_{i-1} / T^{-}_i}{1+\sum_{k=1}^N \Pi^{k}_{i=1} T^{+}_{i-1} / T^{-}_i} & :  j>0.
\end{cases}
\]

We consider only $\psi_j$ for $j >0$
since  $\frac{j}{N} \psi_j$ for $j=0$ has effectively no contribution to the mean abundance of cooperation $\langle X\rangle = \sum_{j=0}^N \frac{j}{N} \psi_j
= \sum_{j=1}^N \frac{j}{N} \psi_j
$.

Let us denote the distribution by $\psi_j  =\psi_{\mathcal{N},\,j}/ \psi_{\mathcal{D}}
$
where the nominator $\psi_{\mathcal{N},\,j}$  and the denominator $\psi_{\mathcal{D}}$ are given by
\begin{align}
\psi_{\mathcal{N},\,j}
&=\Pi^{j}_{i=1}T^{+}_{i-1} /T^{-}_i,
\label{eq_nominator}\\
\psi_{\mathcal{D}} 
&=1+\sum_{k=1}^N \psi_{\mathcal{N},\,k}.
\label{eq_denominator}
\end{align}

To derive the condition for $\langle X \rangle > 1/2$ under weak selection,
we need to compute
\begin{align}
\psi_j \vert_{\beta =0} &= \frac{\psi_{\mathcal{N},\,j}}{\psi_{\mathcal{D}}}  \Big\vert_{\beta =0}, 
\label{eq_dist_0}\\
\pdv{\psi_j}{\beta}  \Big\vert_{\beta =0}
         &=\frac{\psi^{\prime}_{\mathcal{N},\,j} \psi_{\mathcal{D}} -\psi_{\mathcal{N},\,j} \psi^{\prime}_D}{\left(\psi_{\mathcal{D}}\right)^2} \Bigg\vert_{\beta =0},
\label{eq_dist_prime_0}
\end{align}
which are to be inserted into Eq.\,\eqref{eq_mean_abundance_approx}.

From Eqs.\,\eqref{eq_tplus_SDUC} and \eqref{eq_tminus_SDUC},
\begin{align*}
T^+_i &= \frac{N-i}{N} \frac{1}{1 +e^{-\beta\left[ A  -\pi_D(i)\right]}},\\
T^-_i &= \frac{i}{N} \frac{1}{1 +e^{-\beta\left[\pi_C(i) - A \right]}}.
\end{align*}
We then have
\begin{align}
T^+_i |_{\beta=0} & = \frac{N-i}{2N},\label{eq_tplus_0_SDUC}\\
T^-_i |_{\beta=0} & =\frac{i}{2N},\label{eq_tminus_0_SDUC}\\
(T^+_i)'|_{\beta=0} &	=\frac{N-i}{4 N} [A -\pi_D(i)],\\
(T^-_i)'|_{\beta=0} &=\frac{i}{4 N} [\pi_C(i) -A],\label{eq_tminus_prime_0_SDUC}
\end{align}
where $(T^+_i)' =\pdv*{T^+_i}{\beta}$ and $(T^-_i)' =\pdv*{T^-_i}{\beta} $.
Inserting Eqs.\,\eqref{eq_tplus_0_SDUC} and \eqref{eq_tminus_0_SDUC} into Eqs.\,\eqref{eq_nominator} and \eqref{eq_denominator},
we get
\begin{align}
\psi_{\mathcal{N},\,j} |_{\beta=0} &=C^j_N,
\label{eq_nominator_0_SDUC}\\
\psi_{\mathcal{D}} |_{\beta=0} &=2^N,
\label{eq_denominator_0_SDUC}
\end{align}
where $C^j_N =N! /\left(j!(N-j)!\right)$ is a binomial coefficient.
We get
\begin{align}
&\psi^\prime_{\mathcal{N},\,j} \big\vert_{\beta=0} & \nonumber \\
&=\sum_{i =1}^j \left[\frac{\left(T^{+}_{i-1}\right)^\prime T^{-}_i -T^{+}_{i-1} \left(T^{-}_i\right)^\prime}{\left(T^{-}_i\right)^2}  \Pi^{j}_{k=1, k\ne i}\frac{T^{+}_{k-1}}{T^{-}_k}\right] \Bigg\vert_{\beta=0},&
\label{eq_nominator_prime_SDUC}\\
& \psi_{\mathcal{D}}^\prime |_{\beta=0} 
=\sum_{j=1}^N \psi^\prime_{\mathcal{N},\,j} \big\vert_{\beta=0}.
\end{align}

From Eq.\,\eqref{eq_tplus_0_SDUC} to \ref{eq_tminus_prime_0_SDUC},
we get
\begin{align}
& \frac{\left(T^{+}_{i-1}\right)^\prime T^{-}_i -T^{+}_{i-1} \left(T^{-}_i\right)^\prime}{\left(T^{-}_i\right)^2}  \bigg\vert_{\beta=0} & \nonumber \\
&	=\frac{N-(i-1)}{2i} [2A - \pi_C(i) -\pi_D(i-1)], \label{eq_nominator_prime_factor_SDUC} 
\\
& \Pi^{j}_{k=1, k\ne i}\frac{T^{+}_{k-1}}{T^{-}_k} \bigg\vert_{\beta=0}
	=\frac{i}{N-(i-1)} C^j_N.
\label{eq_denominator_prime_factor_SDUC}
\end{align}

Inserting Eqs.\,\eqref{eq_nominator_prime_factor_SDUC} and \eqref{eq_denominator_prime_factor_SDUC} into Eq.\,\eqref{eq_nominator_prime_SDUC},
we get	
\begin{align}
\psi^\prime_{\mathcal{N},\,j} \big\vert_{\beta=0}
&=\frac{C^j_N}{2} \sum_{i =1}^j  \left[2A - \pi_C(i) -\pi_D(i-1)\right],
\label{eq_nominator_prime_0_SDUC} \\
\psi_{\mathcal{D}}^\prime |_{\beta=0}  
&=\sum_{j=1}^N \frac{C^j_N}{2} \sum_{i=1}^j  [2A  -\pi_C(i) -\pi_D(i-1)].
\label{eq_denominator_prime_0_SDUC}
\end{align}	
	
Inserting Eqs.\,\eqref{eq_nominator_0_SDUC}, \eqref{eq_denominator_0_SDUC}, \eqref{eq_nominator_prime_0_SDUC}, and \eqref{eq_denominator_prime_0_SDUC} into
Eqs.\,\eqref{eq_dist_0} and \eqref{eq_dist_prime_0},
we get	
\begin{equation}
\psi_j \big |_{\beta=0} = \frac{C^j_N}{2^N},
\end{equation}
\begin{equation}
\begin{split}
\pdv{\psi_j}{\beta}  \Big\vert_{\beta =0} 
=\frac{C^j_N}{2^{2N+1}} \left\{ A 2^{N+1} j
        -2^N\sum_{i=1}^{j}\left[\pi_C(i) +\pi_D(i-1)\right] \right. \\ 
        \left. - A  N2^N    
       	+\sum_{k=1}^{N}C^{k}_N\sum_{i=1}^{k}\left[\pi_C(i) +\pi_D(i-1)\right] \right\}
\end{split}
\end{equation}
where
$        \pi_C(i) +\pi_D(i-1) 
        = 2(i-1) /(N-1) -\rho 
$
according to Eqs.\,\eqref{eq_payoff_c} and \eqref{eq_payoff_d}.

Using 
\begin{align}
\sum_{j=1}^N j^3 C^j_N &=N^2(N+3)2^{N-3}, \\
\sum_{j=1}^N j^2 C^j_N &=N(N+1)2^{N-2}, \\
\sum_{j=1}^N j C^j_N &=N2^{N-1},
\end{align}
we get
\begin{align}
\sum_{j=1}^N \frac{j}{N} \psi_j \big |_{\beta=0} &=\frac{1}{2}, \\
\sum_{j=1}^N \frac{j}{N}\pdv{\psi_j}{\beta}  \Big\vert_{\beta =0} &=\frac{1}{8} \left(2A +\rho -1\right).
\end{align}
Then we get
\begin{align}
\langle X\rangle 
&\approx \sum_{j=1}^N \frac{j}{N} \psi_j \big |_{\beta=0}  +\sum_{j=1}^N \frac{j}{N}\pdv{\psi_j}{\beta}  \Big\vert_{\beta =0} \nonumber \\
&  =\frac{1}{2} +\frac{1}{8} \left(2A +\rho -1\right) \beta. \tag{\ref{eq_mean_abudance_SDUC}}
\end{align}

\section{Derivation of Eqs.\,\eqref{eq_sduc_mode_1} and \eqref{eq_sduc_mode}\label{deriv_sduc_mode}
}

From Eqs.\,\eqref{eq_tplus_SDUC} and \eqref{eq_tminus_SDUC},
we have
\begin{equation}
T^+_{i-1} /T^-_{i}
=\frac{N-i+1}{i} \frac{1 +e^{-\beta\left[\pi_C(i) - A \right]}}{1 +e^{-\beta\left[ A  -\pi_D(i-1)\right]}}.
\label{eq_tplus_over_tminus}
\end{equation}

Since $T^+_{0} /T^-_1 >1$,
$T^+_{N-1} /T^-_N <1$,
and $T^+_{i^*-1} /T^-_{i^*}$ strictly decreases with real numbers $i^* \in (0,N) \subset \mathbb{R}$,
there is a single real number $i^*$ that satisfies
\begin{equation*}
T^+_{i^*-1} /T^-_{i^*} =1,
\tag{\ref{eq_sduc_mode_1}}
\end{equation*}
which yields
\begin{equation*}
\frac{i^*}{N +1}  =\frac{1 +e^{-\beta\left[\pi_C(i^*) - A \right]}}{2 +e^{-\beta\left[\pi_C(i^*) - A \right]} +e^{-\beta\left[ A  -\pi_D(i^*-1)\right]}}.
\tag{\ref{eq_sduc_mode}}
\end{equation*}
 Note that $i^*$ is a real number approximation of the mode
that is an integer,
the most frequent number of cooperators.
For integers $i \in [0,N] \subset\mathbb{Z}$,
$T^+_{i-1} /T^-_i$ strictly decreases with $i$.
Then we have
$T^+_{i-1} /T^-_i >1$ for $i  <\lfloor i^*\rfloor$, $T^+_{i-1} /T^-_{i} \ge 1$ for $i =\lfloor i^*\rfloor$  and $T^+_{i-1} /T^-_i  <1$ for $i >\lfloor i^*\rfloor$
where $\lfloor i^*\rfloor$ denotes the largest integer that is less than or equal to $i^*$.
Since $\psi_i =\psi_{i-1} T^+_{i-1} /T^-_i$,
the discrete distribution $\psi_i$ picks at $i=\lfloor i^*\rfloor$
that is the (integer) mode of the distribution (Fig.\,\ref{fig_correspondence}).

\section{Derivation of Eq.\,\eqref{eq_dist_approx}
\label{deriv_dist_approx}
}

Under weak selection $\beta \ll 1$,
we have
\begin{align}
&\frac{T^+_{j-1}}{T^-_j} \nonumber \\
&\approx \frac{N -j+1}{j} \left(1 -\frac{\rho\beta}{2}\right) \exp\left[-\beta\left(\frac{j -1}{N-1} -\rho - A \right)\right]. &
\label{eq_transition_prob_approx}
\end{align}

Inserting Eq.\,\eqref{eq_transition_prob_approx} into  Eq.\,\eqref{eq_stat_dist_recurrence}
as well as using $\left(1 -\frac{\rho}{2}\beta\right)^k \approx e^{-k\frac{\rho}{2}\beta}$
and $\frac{N!}{k!(N-k)!}p^k(1-p)^{N -k} \approx \frac{1}{\sqrt{2\pi Np(1-p)}}\exp[-\frac{(k-Np)^2}{2Np(1-p)}]$ \cite{boas2005mathematical},
we get
\begin{equation}
	\psi_i / \psi_0 
	\propto
	 \exp[-\frac{(i -\mu)^2}{2\sigma^2}]
\tag{\ref{eq_dist_approx}}
\end{equation}
where 
\begin{align}
\mu &=\frac{N(N-1)}{4(N-1) +N\beta}\left(2 +\left[A +\frac{\rho}{2}+\frac{1}{2(N-1)}\right]\beta\right), \tag{\ref{eq_mean}}\\
\sigma^2   &=\frac{N(N-1)}{4(N-1) +N\beta}.
\tag{\ref{eq_deviation}}
\end{align}

\bibliographystyle{apsrev4-1}
\bibliography{SDUC_submission_3.bbl} 

\end{document}